\documentclass[useAMS,referee,figuresright,usenatbib]{biom}
\usepackage{fancyhdr} 
\usepackage{latexsym, epsfig, verbatim, amsmath, amssymb, lscape, graphics, threeparttable, color}
\usepackage{mathtools, multirow, rotating}
\usepackage{booktabs}
\pdfoutput=1
\allowdisplaybreaks

\def\bbeta{{\boldsymbol \beta}}
\def\bgamma{{\boldsymbol \gamma}}
\def\bpi{{\boldsymbol \pi}}
\def\bSigma{{\boldsymbol \Sigma}}
\def\btheta{{\boldsymbol \theta}}

\def\bnu{{\boldsymbol \nu}}
\def\bA{{\boldsymbol A}}

\def\bD{{\boldsymbol D}}
\def\boldf{{\boldsymbol f}}
\def\bF{{\boldsymbol F}}
\def\bG{{\boldsymbol G}}
\def\bS{{\boldsymbol S}}
\def\bH{{\boldsymbol H}}

\def\bg{{\boldsymbol g}}

\def\bI{{\boldsymbol I}}
\def\bJ{{\boldsymbol J}}
\def\bK{{\boldsymbol K}}
\def\bl{{\boldsymbol l}}

\def\bt{{\boldsymbol t}}

\def\bV{{\boldsymbol V}}
\def\bW{{\boldsymbol W}}
\def\bX{{\boldsymbol X}}

\def\bZ{{\boldsymbol Z}}
\def\bzero{{\boldsymbol 0}}
\def\tbeta{{\widetilde \beta}}
\def\tbbeta{{\widetilde \bbeta}}
\def\teta{{\widetilde \eta}}
\def\tbpi{{\widetilde \bpi}}
\def\tbtheta{{\widetilde \btheta}}

\def\hbbeta{{\widehat \bbeta}}
\def\hbpi{{\widehat \bpi}}
\def\hbSigma{{\widehat \bSigma}}
\def\hbtheta{{\widehat \btheta}}
\def\hbnu{{\widehat \bnu}}

\def\mathP{{\mathbb P}}
\def\mathR{{\mathbb R}}

\def\dbg{{\dot{\boldsymbol g}}}

\def\dbm{{\dot{\boldsymbol m}}}

\def\ddbm{{\ddot{\boldsymbol m}}}
\def\tbg{{\tilde{\boldsymbol g}}}

\def\tbm{{\tilde{\boldsymbol m}}}

\begin{document}

\title{Integration of Summary Information from External Studies for Semiparametric Models}

\author{Jianxuan Zang$^{1,*}$\email{jxzang00@uw.edu}, 
K.C.G. Chan$^{2,**}$\email{kcgchan@u.washington.edu}, and 
Fei Gao$^{3,***}$\email{fgao@fredhutch.org} \\
$^{1}$Department of Statistics, University of Washington, Seattle, Washington, U.S.A.  \\
$^{2}$Department of Biostatistics, University of Washington, Seattle, Washington, U.S.A.  \\
$^{3}$Vaccine and Infectious Disease Division, Fred Hutchinson Cancer Center, Seattle, Washington, U.S.A.}

\begin{abstract}
With the development of biomedical science, researchers have increasing access to an abundance of studies focusing on similar research questions. There is a growing interest in the integration of summary information from those studies to enhance the efficiency of estimation in their own internal studies. In this work, we present a comprehensive framework on integration of summary information from external studies when the data are modeled by semiparametric models.
Our novel framework offers straightforward estimators that update conventional estimations with auxiliary information. It addresses computational challenges by capitalizing on the intricate mathematical structure inherent to the problem. We demonstrate the conditions when the proposed estimators are theoretically more efficient than initial estimate based solely on internal data. Several special cases such as proportional hazards model in survival analysis are provided with numerical examples.
\end{abstract}

\maketitle
\section{Introduction}

With the development of biomedical research, researchers are progressively exploring the integration of information derived from external studies to enhance the efficacy of statistical inference in their own study.
Often, access to individual-level data is restricted due to privacy concerns, leaving summary information from external studies as the primary resource.
To harness this external knowledge effectively, the development of a comprehensive statistical inference framework is essential.

Recent studies in this domain have primarily focused on the incorporation of auxiliary information from population-based data sources, such as census data or healthcare databases.
Several frameworks have emerged, including frequentist methods (e.g., \cite{qin2000miscellanea}; \cite{chatterjee2016constrained}; \cite{han2019empirical}, \cite{gao2023noniterative}) and Bayesian methods (e.g., \cite{cheng2019informing}; \cite{boonstra2020incorporating}).

However, when auxiliary information is derived from external studies with limited sample sizes, the inherent uncertainty in summary information from external studies must be accounted for.
Researchers have made noteworthy efforts in this direction.
For instance, \cite{kundu2019generalized} developed a generalized meta-analysis approach for multivariate regression model based on the generalized method of moments approach, allowing the combination of information across multiple studies while accounting for parameter estimate uncertainties.
Motivated by the empirical likelihood approach, \cite{zhang2020generalized} proposed a generalized integration model approach to combine individual data with summary information with uncertainty, which was later extended to data arise from case-control studies \citep{zhang2021integrative}.
Furthermore, \cite{huang2020unified} explored the analysis of right-censored data when additional information on the regression coefficients evaluated in a reduced Cox model is available.
They employed a generalized method of moments approach incorporating uncertainty of the external information in the inference procedure.

Despite these valuable contributions, a comprehensive estimation framework that offers a straightforward, noniterative update procedure for incorporating such auxiliary data remains conspicuously absent.
Additionally, existing studies primarily concentrates on parametric models and often neglect semiparametric models characterized by additional infinite dimensional parameters.
While specific instances of semiparametric models have been examined in works such as \cite{huang2020unified} and \cite{han2022semiparametric}, a unified framework applicable to the entire spectrum of semiparametric models is lacking.

In this study, we bridge this gap by extending the general framework of \cite{gao2023noniterative}, which was proposed to incorporate the population-based auxiliary information where variability can be neglected.
We capitalize on the intricate mathematical architecture inherent to the problem to obviate the need for iterative algorithm, which highly improves the computing efficiency. 
Furthermore, our framework accommodates scenarios in the presence of multiple external studies.
We provide the details of applying our proposed method in various commonly used parametric and semiparametrc models. Simulation experiments are conducted to demonstrate the effectiveness of our approach. Finally, we illustrate an application of our method by evaluating the cancer risk using data from Women’s Health Initiative. 

The rest of the manuscript is organized as follows. In Section 2, we introduce the model setting and construct the general framework of incorporating auxiliary information with variability. In Section 3, we propose the simulation studies including parametric and semiparametric models. In Section 4, we apply our framework on the WHI dataset. Finally, in Section 5 we conclude the paper with a brief summary and discuss the future work.

\section{Methods}

\subsection{Model and Data}
Let $\bX_i$ $(i = 1,\dots, n)$ be \textit{i.i.d} observations of a random variable $\bX$. We assume that the distribution of $\bX$ is associated with an unknown $p$-dimensional  parameter $\btheta\in\Theta\subset\mathR^p$ that is of primary interest and an infinite dimensional nuisance parameter $\eta$.
Suppose that the true $(\btheta_0,\eta_0)$ of $(\btheta,\eta)$ maximizes a criterion function $\mathP\left\{m(\bX;\btheta,\eta)\right\}$, where $\mathP$ is the probability measure with respect to $\bX$.
We may construct an estimator $(\tbtheta,\teta)$ by maximizing the empirical criterion function
\[(\tbtheta,\teta) = \text{arg max}_{\btheta,\eta}\mathP_n m(\bX;\btheta,\eta),\]
where $\mathP_n$ denote the empirical measure.
In the special case of maximum likelihood estimator, the function $m(\bX;\btheta,\eta)$ is taken to be the log-likelihood function $\log f(\bX;\btheta,\eta)$, where $f(\bX;\btheta,\eta)$ is the density of $\bX$.

We first consider the case incorporating summary information from one external study, while extension to multiple external studies is given in Section \ref{sec:mult_ext}.
Suppose that some information on the distribution of $\bX$ can be obtained from an external study, such that an estimate $\tbbeta$ of an $r$-dimensional parameter $\bbeta$ is known.
We assume that some linkage between the internal and external models is known such that $\bbeta$ satisfies $\mathP\{\bg(\bX;\btheta,\eta,\bpi,\bbeta)\} = \bzero$, where $\bg(\cdot)$ is a $q$-vector function and $\bpi$ is a $v$-vector parameter whose estimate is not available from the external study.
We assume that data from the external study is unavailable, however, the sample size $N$ of the external study is known. Specifically, assume 
\begin{assumption}
    $N/n \rightarrow \rho \in (0, \infty)$ as $n \rightarrow \infty$.
\end{assumption}

\begin{assumption}
    $\sqrt N (\tbbeta - \bbeta_0) = U + o_{\tilde \mathP}(1)$, where $U\sim N(0,\bSigma_0)$ and $\tilde \mathP$ is the probability measures in the external study, and $\bSigma_0$ is the covariance matrix of $\tbbeta$.
\end{assumption}

\begin{remark}
Here, we consider a general framework where the external study may not share the same set of parameters with the internal study.
In addition, there may be other parameters $\bpi$ in the model of the external study, whose estimate may not be available to the researcher.
In the special case where estimates for one or more components of $\btheta$ is available from the external study, we may additionally include constant function in $\bg$, e.g., setting $g_1(\bX;\btheta,\eta,\bpi,\bbeta) = \theta_1-\beta_1$ to indicate that an estimate of the first element of $\btheta$ is available from the external study.
\end{remark}

The proposed framework is general and includes many commonly used models and estimation approaches.
\begin{example}[Logistic Regression Model]
Suppose that we observe an i.i.d. sample of $\bX \equiv (Y,Z)$. We consider the logistic regression for the binary outcome $Y$ in our internal study, where 
$\text{logit}\left\{\Pr(Y=1|Z_1,Z_2)\right\} = \theta_0+Z_1\theta_1+Z_2\theta_2+\epsilon$ and $\text{logit}(\cdot)$ is the logit function.
Suppose that the estimate coefficient of reduced model $\tilde\beta_1$ is given as our auxiliary information, where  $\text{logit}\left\{\Pr(Y=1|Z_1,Z_2)\right\}=\alpha_1+Z_1\beta_1$.
The information from external studies can be summarized by
$g(\bX;\theta,\alpha,\beta) =\{\text{expit}(\alpha_1+\beta_1Z_1)-\text{expit}(\theta_0+\theta_1Z_1+\theta_2Z_2)\}(1, Z_1)^{\rm T}$, where $\text{expit}(\cdot)$ is the expit function.

\end{example}

\begin{example}[Proportional Hazards Model and $t$-Year Survival Probability]
Suppose that the internal study concerns the regression analysis of censored survival time $(Y,\Delta)$ on a binary treatment $Z$ using the proportional hazards model, such that $\bX = (Y,\Delta,Z)$, $m(\theta,\lambda) = \log\Lambda\{Y\} + \theta Z - \int_0^Y\exp(\theta Z)\lambda(t)dt$ is the nonparametric log-likelihood function, where $T$ and $C$ are event time and censoring time, $Y=\min(T,C)$, $\Delta = I(T\le C)$, $\theta$ is the regression coefficient, $\lambda(\cdot)$ is the baseline hazard function, and $\Lambda\{u\}$ is the jump size of $\lambda$ at $u$.
Suppose that a $t$-year survival probability for those receive $Z=0$ is known from an external study, i.e., $\beta = \Pr(T\ge t|Z=0)$.
Then, the information from the external study can be summarized by $g(\bX;\theta,\lambda,\beta) = \exp\{-\int_0^t\lambda(u)du\} - \beta$.
\end{example}

Similar to \cite{zhang2020generalized}, an updated estimator for $\btheta$ that incoporates auxiliary information can be obtained by the empirical likelihood approach.
Let $p_i$ be a point mass corresponding to subject $i$ in the internal study.
The empirical likelihood estimator maximizes the joint log-likelihood function 
\[\sum_{i=1}^n\{m(\bX_i; \btheta,\eta) + \log p_i\} - \frac{N}{2}(\tbbeta - \bbeta)^{\rm T}\bV^{-1}(\tbbeta - \bbeta),\]
subject to the constraints $p_i\ge 0$ for $i=1,\dots,n$, $\sum_{i=1}^np_i=1$, and $\sum_{i=1}^np_ig(\bX_i;\btheta,\eta,\bpi,\bbeta) = 0$.
Here, $\bV$ is a given $r\times r$ positive definite matrix which will be discussed further later.

By applying the Lagrange multiplier arguments, it can be seen that the empirical likelihood estimator maximizes
\begin{equation}
\sum_{i=1}^nm(\bX_i;\btheta,\teta)-\log\left\{1+\bt^{\rm T}\bg(\bX_i;\btheta,\eta,\bpi,\tbbeta)\right\}- \frac{N}{2}(\tbbeta - \bbeta)^{\rm T}\bV^{-1}(\tbbeta - \bbeta),\label{equ:lik}
\end{equation}
with $\bt$ satisfies
\[\sum_{i=1}^n\frac{\bg(\bX_i;\btheta,\eta,\bpi,\tbbeta)}{1+\bt^{\rm T}\bg(\bX_i;\btheta,\eta,\bpi,\tbbeta)}=\bzero.\]

\begin{remark}
The objective function is similar to that in \cite{zhang2020generalized}, which considered the integration of external summary in a regression model framework.
Here, our framework is more general in three aspects.
First, our framework does not require a ``full'' regression model in the internal study and a reduced or misspecified working model in the external study, as in \cite{zhang2020generalized}.
Second, we incorporate the setting when the internal study is modeled through a semiparametric model, which poses additional theoretical and numerical challenges.
Third, our framework obviates the need for iterative algorithms to attain optimal efficiency. Instead, we circumvent the computational overhead by harnessing the intricate mathematical structure inherent to the problem.
\end{remark}

To formulate an update formula, we will leverage the asymptotic distributions of the estimators with and without incorporating auxiliary information. We expect to exploit a special structure that the estimator that incorporates additional summary information can be expressed as the sum of the initial estimator and a matrix factor, with a residual term that is asymptotically negligible.

Let $\dot{\boldf}_\bgamma(\cdot)$ be the generic notation for the derivative of a function $\boldf$ with respective to a finite-dimensional parameter $\bgamma$.
By applying Taylor expansion to the objective function (\ref{equ:lik}) (more details given in Web Appendix A.1), we show that under some regularity conditions, the asymptotic distribution of the updated estimator $(\hbtheta,\hbpi,\hbbeta)$ is given by
\begin{align}
\sqrt n\begin{pmatrix}\hbtheta-\btheta_0\\\hbpi - \bpi_0\\\hbbeta - \bbeta_0\end{pmatrix} &= \begin{pmatrix}\bI_{(p+v)\times(p+v)}&\bzero_{(p+v)\times q}&\bzero_{(p+v)\times r}&\bzero_{(p+v)\times r}\\\bzero_{r\times (p+v)}&\bzero_{r\times q}&\bI_{r\times r}&\bzero_{r\times r}\end{pmatrix}\nonumber\\
&\times\bA(\btheta_0,\eta_0,\bpi_0,\bbeta_0)^{-1}\bl(\btheta_0,\eta_0,\bpi_0,\bbeta_0)+ o_\mathP(1),\label{equ:para_func_final}
\end{align}
where 

{\fontsize{10}{12}\selectfont$
\bA(\btheta,\eta,\bpi,\bbeta) = \begin{pmatrix}-\mathP\tbm_{\btheta\btheta}(\bX;\btheta,\eta) &\bzero_{p\times v}& \mathP\tbg_\btheta^{\rm T}(\bX;\btheta,\eta,\bpi,\bbeta)&\bzero_{p\times r}&\bzero_{p\times r}\\
\bzero_{v\times p}&\bzero_{v\times v}&\mathP\dbg_\bpi^{\rm T}(\bX;\btheta,\eta,\bpi,\bbeta)&\bzero_{v\times r}&\bzero_{v\times r}\\
-\mathP\tbg_\btheta(\bX;\btheta,\eta,\bpi,\bbeta)&- \mathP\dbg_\bpi (\bX;\btheta,\eta,\bpi,\bbeta)&\mathP \bG(\bX;\btheta,\eta,\bpi,\bbeta)&- \mathP\dbg_\bbeta(\bX;\btheta,\eta,\bpi,\bbeta)&\bzero_{q\times r} \\
\bzero_{r\times p}&\bzero_{r\times v}&\mathP\dbg_\bbeta^{\rm T}(\bX;\btheta,\eta,\bpi,\bbeta)&\rho \bV^{-1}&-\rho \bV^{-1}\\
\bzero_{r\times p}& \bzero_{r\times v} &  \bzero_{r\times q} & \bzero_{r\times r}& \bI_{r\times r}\end{pmatrix}$,}
\[\bl(\btheta,\eta,\bpi,\bbeta) = \left(\sqrt n\mathP_n \tbm_\btheta(\bX;\btheta,\eta)^{\rm T}, \bzero_{v\times1}^{\rm T},\sqrt n\mathP_n \tbg(\bX;\btheta,\eta,\bpi,\bbeta)^{\rm T},\bzero_{r\times1}^{\rm T}, \sqrt \rho U^{\rm T}\right)^{\rm T},\]
and the functions $\tbm_{\btheta\btheta}(\bX;\btheta,\eta),\tbg_\btheta(\bX;\btheta,\eta,\bpi,\bbeta),\tbm_\btheta(\bX;\btheta,\eta), \tbg(\bX;\btheta,\eta,\bpi,\bbeta)$, and $\bG(\bX;\btheta,\eta,\bpi,\bbeta)$ are versions of $\ddbm_{\btheta\btheta}(\bX;\btheta,\eta),\dbg_\btheta(\bX;\btheta,\eta,\bpi,\bbeta),\dbm_\btheta(\bX;\btheta,\eta), \bg(\bX;\btheta,\eta,\bpi,\bbeta)$, and $\bg^{\otimes2}(\bX;\btheta,\eta,\bpi,\bbeta)$ with the nonparametric component $\eta$ profiled out that, and they are formally defined in Web Appendix A.1.

To provide an updating formula for $(\btheta,\bpi,\bbeta)$, we would need to obtain initial estimator for $\bpi$, which may not be available form the external studies. 
Specifically, we can obtain the initial estimate $\tbpi$ by solving $\sum_{i=1}^n\bg^*(\bX_i;\tbtheta,\teta,\bpi,\tbbeta) = \bzero$, where $\bg^*$ can without loss of generality be defined as the first $v$-element of $\bg$. The specific choice of $\bg^*$ does not impact the estimation of $\tbtheta$, as discussed in detail in \cite{gao2023noniterative}.

Following a similar Taylor expansion, the asymptotic distribution of the initial estimator $(\tbtheta,\tbpi,\tbbeta)$ is given by 
\[\resizebox{\hsize}{!}{$
\sqrt n\begin{pmatrix}\tbtheta-\btheta_0\\\tbpi - \bpi_0\\\tbbeta - \bbeta_0\end{pmatrix} 
=\begin{pmatrix}-\mathP\tbm_{\btheta\btheta} (\bX;\btheta_0,\eta_0)&\bzero_{p\times v}&\bzero_{p\times r}\\
-\mathP\tbg^*_{\btheta}(\bX;\btheta_0,\eta_0,\bpi_0,\bbeta_0)&-\mathP\dbg^*_{\bpi}(\bX;\btheta_0,\eta_0,\bpi_0,\bbeta_0) & -\mathP\dbg^*_{\bbeta}(\bX;\btheta_0,\eta_0,\bpi_0,\bbeta_0)\\
\bzero_{r\times p}&\bzero_{r\times v}&-\bI_{r\times r}
\end{pmatrix}^{-1}$}\]

{\fontsize{10}{12}\selectfont$\times \begin{pmatrix}
-\mathP\tbm_{\btheta\btheta}(\bX;\btheta_0,\eta_0) & \bzero_{p\times v} & \mathP\tbg_\btheta^{\rm T}(\bX;\btheta_0,\eta_0,\bpi_0,\bbeta_0) &  \bzero_{p\times r}& \bzero_{p\times r} \\
-\mathP\tbg^*_{\btheta}(\bX;\btheta_0,\eta_0,\bpi_0,\bbeta_0) & -\mathP\dbg^*_{\bpi}(\bX;\btheta_0,\eta_0,\bpi_0,\bbeta_0) & \mathP\bG^*(\bX;\btheta_0,\eta_0,\bpi_0,\bbeta_0)&-\mathP\dbg^*_{\bbeta}(\bX;\btheta_0,\eta_0,\bpi_0,\bbeta_0)&\bzero_{v\times r} \\
\bzero_{r\times p}& \bzero_{r\times v}& \bzero_{r\times v}  & \bzero_{r\times r} & -\bI_{r\times r}
\end{pmatrix}$}

\begin{align}
\times \bA(\btheta_0,\eta_0,\bpi_0,\bbeta_0)^{-1}\bl(\btheta_0,\eta_0,\bpi_0,\bbeta_0)+ o_\mathP(1),\label{equ:original}
\end{align}

where $\bG^* = (\bI_{v\times v} \ \bzero_{v\times (q-v)})\bG$.

Taking the difference between (\ref{equ:para_func_final}) and (\ref{equ:original}), we obtain that
\begin{align}
\begin{pmatrix}\hbtheta\\\hbpi\\\hbbeta\end{pmatrix}
&= \begin{pmatrix}\tbtheta\\\tbpi\\\tbbeta \end{pmatrix} + n^{-1/2}\begin{pmatrix}
\bS_\btheta(\btheta,\eta,\bpi,\bbeta)\\
\bS_\bpi(\btheta,\eta,\bpi,\bbeta)\\
\bS_\bbeta(\btheta,\eta,\bpi,\bbeta)
\end{pmatrix}\bA(\btheta_0,\eta_0,\bpi_0,\bbeta_0)^{-1}\bl(\btheta_0,\eta_0,\bpi_0,\bbeta_0) + o_\mathP(n^{-1/2}),\label{equ:update}
\end{align}
where \begin{align*}
\bS_\btheta(\btheta,\eta,\bpi,\bbeta) =& \begin{pmatrix}
\bzero_{p\times p} &\bzero_{p\times v}& \mathP \tbm_{\btheta\btheta}(\bX_i;\btheta,\eta)^{-1} \mathP\tbg_\btheta(\bX_i;\btheta,\eta,\bpi,\bbeta)^{\rm T} &\bzero_{p\times r}&\bzero_{p\times r}\end{pmatrix},\\
\bS_\bpi(\btheta,\eta,\bpi,\bbeta) =& -\mathP\dbg^*_{\bpi}(\bX_i;\btheta,\eta)^{-1}\begin{pmatrix}
\bzero_{v\times p} & \bzero_{v\times v} & \bH & \mathP\dbg^*_{\bbeta}(\bX_i;\btheta,\eta,\bpi,\bbeta)&-\mathP\dbg^*_{\bbeta}(\bX_i;\btheta,\eta,\bpi,\bbeta)\end{pmatrix},\\
\bS_\bbeta(\btheta,\eta,\bpi,\bbeta) =&\begin{pmatrix}
\bzero_{r\times p} & \bzero_{r\times v} &\bzero_{r\times q} & \bI_{r\times r}&-\bI_{r\times r}\end{pmatrix},
 \end{align*}
 and
$
\bH = \mathP\tbg^*_{\theta}(\bX_i;\btheta,\eta,\bpi,\bbeta)\mathP\tbm_{\btheta\btheta}(\bX_i;\btheta,\eta)^{-1} \mathP\tbg_\btheta(\bX_i;\btheta,\eta,\bpi,\bbeta)^{\rm T}-\mathP\bG^*(\bX_i;\btheta,\eta,\bpi,\bbeta)$.

To this end, we propose a one-step updating formula based on equation (\ref{equ:update}) by replacing the true values $(\btheta_0,\eta_0,\bpi_0,\bbeta_0)$ by the initial estimators $(\tbtheta,\teta,\tbpi,\tbbeta)$, replacing the probability measure by the empirical measure $\mathP_n$, and neglecting the residual $o_\mathP(1)$ terms.
With a slight abuse of notation, we denote (\ref{equ:update}) by $(\hbtheta,\hbpi,\hbbeta)$ because they are asymptotically equivalent. 
Theorem \ref{thm:1} shows that the resulting estimator $(\hbtheta,\hbpi,\hbbeta)$ is consistent for the parameters $(\btheta_0,\bpi_0,\bbeta_0)$.
\begin{theorem}\label{thm:1}
Let $\bnu = (\btheta,\bpi,\bbeta),\Omega_n = \left\{\bnu:\left\|\bnu-\bnu_0\right\| \leq n^{-1/3}\right\}$. Under Assumptions 1-2 and some regularity conditions given in the Web Appendix A.2, the joint log-likelihood function attains its maximum value at $\bnu=\hbnu$ with probability 1 such that the first partial derivative is zero and $\hbnu$ is in the interior of the closed ball $\Omega_n$.
\end{theorem}

From the asymptotic distribution of $(\hbtheta,\hbpi,\hbbeta)$ given in (\ref{equ:para_func_final}), the variance of $(\hbtheta,\hbpi,\hbbeta)$ is given by $\bD\bW\bD^{\rm T}$,
where 
\[\bD =\begin{pmatrix}\bI_{(p+v)\times(p+v)}&\bzero_{(p+v)\times q}&\bzero_{(p+v)\times r}&\bzero_{(p+v)\times r}\\\bzero_{r\times (p+v)}&\bzero_{r\times q}&\bI_{r\times r}&\bzero_{r\times r}\end{pmatrix}\bA(\btheta_0,\eta_0,\bpi_0,\bbeta_0)^{-1}\]
and
\[\resizebox{\hsize}{!}{$
\bW  = \begin{pmatrix}\mathP\tbm_\btheta^{\otimes2}(\bX;\btheta_0,\eta_0)&\bzero_{p\times v}&\mathP\tbm_\btheta(\bX;\btheta_0,\eta_0)\tbg^{\rm T}(\bX;\btheta_0,\eta_0,\bpi_0,\bbeta_0)&\bzero_{p\times q}&\bzero_{p\times r}\\
\bzero_{v\times p}&\bzero_{v\times v}& \bzero_{v\times q}&\bzero_{v\times r}&\bzero_{r\times r}\\
\mathP\tbg(\bX;\btheta_0,\eta_0,\bpi_0,\bbeta_0)\tbm_\btheta^{\rm T}(\bX;\btheta_0,\eta_0)&\bzero_{q\times v}&\mathP \tbg^{\otimes2}(\bX;\btheta_0,\eta_0,\bpi_0,\bbeta_0)
&\bzero_{q\times r}&\bzero_{q\times r}\\
\bzero_{r\times p}&\bzero_{r\times v}& \bzero_{r\times q}&\bzero_{r\times r}&\bzero_{r\times r}\\
\bzero_{r\times r}&\bzero_{r\times v}&\bzero_{r\times q}&\bzero_{r\times r}&\frac{1}{\rho}\Sigma_0
\end{pmatrix}
$}\]

To estimate the covariance matrix of $(\hbtheta,\hbpi,\hbbeta)$, we would like to replace the true values $(\btheta_0,\eta_0,\bpi_0,\bbeta_0)$ by the initial (or updated) estimators $(\tbtheta,\teta,\tbpi,\tbbeta)$ and replacing the probability measures by the empirical measures.
Specifically, estimation of $\bW$ would involve estimation of $\bSigma_0$, which may not be available from the external study.
In this case, we can use the sandwich estimator denoted in \cite{white1982maximum} from our internal studies. 
Let $\tilde \mathP_N$ denote the empirical measures in the external study. Assume that the summary information $\tbbeta$ in the external study is solved through $\tilde\mathP_N\bF(X;\bbeta) = \bzero$. The sandwich estimator of covariance matrix $\bSigma_0$ is
\[\widehat{\bSigma}_0 = \widehat{\bJ}^{-1}\widehat{\bK}\widehat{\bJ}\]
where $\widehat{\bJ} = \tilde\mathP_N \bF_{\bbeta}(\bX;\tbbeta)$ and $\widehat{\bK} = \tilde\mathP_N \bF(\bX;\tbbeta)\bF(\bX;\tbbeta)^{\rm T}$. Since individual-level data from external studies are not available, we get the empirical estimator from the internal study, based on the assumption that the internal and external studies share the same probability measure.

In the Web Appendix A.3, we demonstrate that  when the initial estimator is from the semiparametric maximum likelihood framework and under some regularity conditions, the asymptotic variance of $\sqrt n (\hbtheta-\btheta_0)$  is smaller than $\sqrt n (\tbtheta- \btheta_0)$ and is minimized  when $\bV = \bSigma_0$.
To this end, one may replace $\bV$ in (\ref{equ:update}) by the estimate $\hbSigma_0$ to obtain the most efficient updated estimator.

\subsection{Incorporating Multiple External Studies}\label{sec:mult_ext}
Sometimes, there are more than one available external study. It will be useful if we can incorporate all relevant external information to gain more efficiency.
For external studies $m=1,\dots,M$, let $\tbbeta_m$ be an estimate of a parameter $\bbeta_m$ from the external study $m$.
Then, we may consider the empirical likelihood approach that maximizes the joint log-likelihood function 
\begin{equation}
    \sum_{i=1}^n\{m(\bX_i; \btheta,\eta) + \log p_i\} - \sum_{m=1}^M N_m(\tbbeta_m - \bbeta_m)^{\rm T}\bV_m^{-1}(\tbbeta_m - \bbeta_m)/2,\label{eq:mult_study}
\end{equation}
subject to the constraints $p_i\ge 0$ for $i=1,\dots,n$, $\sum_{i=1}^np_i=1$, and $\sum_{i=1}^np_i\bg_m(\bX_i;\btheta,\eta,\bpi_m,\tbbeta_m) = 0$ for $m=1,\dots,M$.
The maximization can be in principle conducted through the same one-step updating algorithm with $\tbbeta = (\tbbeta_1^{\rm T},\dots,\tbbeta_M^{\rm T})^{\rm T}$, $\bg = (\bg_1^{\rm T},\dots,\bg_M^{\rm T})^{\rm T}$, $\rho = \sum_{m=1}^M N_m/n$, $\bV = Diag (c_1\bV_1,\dots,c_M\bV_M)$,  and $c_m = N_m/\sum_{m=1}^MN_m$, where $Diag(\cdot)$ is the block diagonal matrix.

Note that the joint likelihood function (\ref{eq:mult_study}) pertain to the case when $M$ external studies estimate different parameters $\bbeta_1,\dots,\bbeta_M$.
Indeed, the form of the joint likelihood function would differ if, for example, the external studies give estimates for the same parameter.
Specifically, if estimates for the same parameter $\bbeta$ from $M$ external studies are available, the maximum empirical likelihood estimator should maximize
\begin{equation}
    \sum_{i=1}^n\{m(\bX_i; \btheta,\eta) + \log p_i\} - \sum_{m=1}^M N_m(\tbbeta_m - \bbeta)^{\rm T}\bV_m^{-1}(\tbbeta_m - \bbeta)/2,\label{eq:mult_true}
\end{equation}
subject to the constraints $p_i\ge 0$ for $i=1,\dots,n$, $\sum_{i=1}^np_i=1$, and $\sum_{i=1}^np_i\bg_m(\bX_i;\btheta,\eta,\bpi_m,\tbbeta_m) = 0$ for $m=1,\dots,M$.
The form is different from (\ref{eq:mult_study}).
However, we show in Web Appendix A.4 that the estimator for $\btheta$ maximizing (\ref{eq:mult_study}) indeed maximizes the desired empirical likelihood function (\ref{eq:mult_true}).
That is, to incorporate multiple external studies with estimates on potentially overlapping parameters, we could treat them as if they are for non-overlapping parameters and apply the proposed non-iterative algorithm to obtain updated estimators based on (\ref{eq:mult_study}).

\section{Simulation Studies}

In this section, we perform simulation studies to evaluate the effectiveness and robustness of our proposed methods across various common parametric and semiparametric scenarios. Our objective is to establish the advantages of our methodology through comparative analysis with existing approaches.

\subsection{Parametric model}

First, we address a scenario of a parametric regression model with auxiliary information on the coefficients of reduced models, as described in \cite{zhang2020generalized}. We assume that we observe an i.i.d. sample of $\bX \equiv (Y,Z)$.
The continuous response variable $Y$ follows a linear regression model given by $Y = \theta_0+Z_1\theta_1+Z_2\theta_2+\epsilon$, with $\epsilon\sim N(0,1)$ and $\theta_{00} \equiv (\theta_0, \theta_1,\theta_2)=(0.1,0.1,0.2)$. The covariates $Z_1\sim N(0,1),Z_2\sim N(0,2)$ with $\text{Cov}(Z_1,Z_2)=0.6$.
In the  external study, two reduced models were employed for estimation: $\text{E}(Y|Z_1)=\pi_1+Z_1\beta_1$ and $\text{E}(Y|Z_2)=\pi_2+Z_2\beta_2$, with estimates $\tbeta_1$ and $\tbeta_2$. We assumed that either $\tbeta_2$ or $(\tbeta_1,\tbeta_2)$ were given as auxiliary information.
The internal data is with sample size $n=1000$, and we evaluate the performance of the estimators that incorporate external study with sample size $N=500$, $1000$, and $2000$.

As shown in Web Appendix B, the estimating equations corresponding to the auxiliary information can be summarised as:
\[g(\bX;\btheta,\bpi,\bbeta) =\left(\begin{matrix}
\theta_0 + \theta_1Z_1+\theta_2Z_2- \pi_1-\beta_1Z_{1}\\
\theta_0 + \theta_1Z_1+\theta_2Z_2- \pi_2-\beta_2Z_{2}\\
(\theta_0 + \theta_1Z_1+\theta_2Z_2- \pi_1-\beta_1Z_{1})Z_1 \\
(\theta_0 + \theta_1Z_1+\theta_2Z_2- \pi_2-\beta_2Z_{2})Z_2
\end{matrix}\right)\]
As for the covariance matrix $\bV$, we calculate the sandwich estimator from the internal study as we discussed before.

\begin{table}[p]
\centering\caption{Simulation Results for Parametric Model}
\label{tab:para1}
\begin{tabular}{cccccccccccccc}

\hline

	&		&	\multicolumn{3}{c}{MLE}			&&	\multicolumn{3}{c}{Proposed (given $\tilde\beta_2$)}	&&\multicolumn{4}{c}{Proposed (given $(\tilde\beta_1,\tilde\beta_2$)}								\\\cline{3-4}\cline{6-9}\cline{11-14}
Setting	&		&	Bias	&	SE 	&&	Bias	&	SE	&	CP&RE	&&	Bias	&	SE	&CP  &    RE	\\
N=500	&	$\theta_1$	&	-0.0005	&	0.0349	&&	-0.0005	&	0.0349	&	0.951&1.0&&	0.0001	&	0.0289	&	0.949&1.2	\\
	&	$\theta_2$	&	0.0001	&	0.0245 	&&	0.0001	&	0.0209	&	0.952&1.2	&&	-0.000	&	0.020	&	0.950&1.2\\
N=1000	&	$\theta_1$	&	-0.0003	&	0.0349  	&&	-0.0003	&	0.0349	&	0.949&1.0&&	-0.000	&	0.0253	&	0.951&1.4\\
	&	$\theta_2$	&	-0.0001 &	0.0247   	&&	-0.000	&	0.0191	&	0.951	&1.3&&	-0.000	&	0.0176	&	0.950&1.4	\\
N=2000	&	$\theta_1$	&	0.0001 	&	0.0350   &&	0.0001	&	0.0350	&	0.946&1.0&&	-0.0003	&	0.0215	&	0.949&1.6	\\
	&	$\theta_2$	&	0.0001	&	0.0247 	&&	0.0000	&	0.0168	&0.949&1.5&&	0.0002	&	0.0145	&0.947&1.7\\
 
\hline

\end{tabular}
\end{table}

Table \ref{tab:para1} show the 
empirical bias (Bias), standard deviation (SE) and coverage probabilities of 95\% confidence intervals (CP) based on 10,000 replications.
It is obvious that upon the incorporation of auxiliary information, our estimator consistently exhibits superior efficiency when compared with the maximum likelihood estimate, which is consistent with the theoretical findings. In addition, comparing the proposed estimator incorporating $(\tbeta_1,\tbeta_2)$ with that incorporating $\tbeta_2$, the variance of the estimator becomes smaller as additional auxiliary information becomes available. For example, for $N=1000$ with only $\tilde\beta_2$ accessible, the variance of our proposed estimator is 1.91; and it reduces to 1.76 when both $(\tilde\beta_1,\tilde\beta_2)$ are available. 
When only $\tilde\beta_2$ is available, the variance of $\theta_1$ remains the same.
This suggests that summary information from a nested working model doesn't enhance inference for omitted covariates in generalized linear model, as discussed in \cite{zhang2020generalized}. Finally, as expected, when the sample size of external data expands, the efficiency of our estimator correspondingly increases. This result is intuitively consistent: with external studies having significantly more samples than the internal study, the variability of $\bbeta$ can be effectively minimized, leading to optimal efficiency.

We have conducted a comparative analysis of the computational speed between our proposed algorithm and the generalized integration model as presented in \cite{zhang2020generalized}, with results given in Table S1 of the Web Appendix B. Remarkably, our method exhibits a computational speed that is in average 60 times faster than the latter. These findings underscore the considerable computational efficiency achieved through our approach when contrasted with the existing methodology. 

We also compare another simulation setting with a binary outcome Y modelled using logistic regression.
The results are shown in Web Appendix B, with a similar performance of the proposed estimators.

\subsection{Survival regression models with t-year survival constraints}

Survival probability data for various cancer sites and subgroups is relatively accessible through the publications of cancer registries. In their work, \cite{huang2016efficient} introduced a distinct structure for the proportional hazards model applied to auxiliary information. Nonetheless, we should notice the limitation of their methodology, particularly in its inability to extend to the broader context of general semiparametric settings and its incapacity to incorporate the variability  of auxiliary information. In this section, we illustrate the application of our method to incorporating subgroup-specific $t$-year survival probabilities to the analysis of right-censored data, adhering to the proportional hazards model.

Let $T$ denote the survival time that follows the proportional hazards model. The cumulative hazard function for $T$ is given as
\begin{align}
    \Lambda(t|\bZ) = \Lambda(t)\exp(\btheta^{\rm T}\bZ),\label{equ:harzard_func}
\end{align}
where $\Lambda(\cdot)$ is an unspecified nondecreasing function, and $\btheta$ is a $p$-dimensional vector of regression parameters. 
Let $C$ denote a censoring time that is conditionally independent of $T$ given the covariates $\bZ$. 
Consequently, we observe $Y \equiv \min (T, C)$ and $\Delta \equiv I(T \leq C)$.  For a random sample comprising $n$ subjects, the observed data consists of $\boldsymbol{X}_i \equiv\left\{Y_i, \Delta_i, \boldsymbol{Z}_i\right\}$ for $i=1, \ldots, n$. The nonparametric maximum likelihood estimator can be readily obtained from the R package {\tt survival}.

Consider that we have acquired the $t_k$-year survival probability for the $k$th subgroup of subjects as auxiliary information for $k=1,\dots,K$.
We denote $\Omega_k$ as the set of covariate values for subjects belonging to subgroup $k$, and $\beta_k$ represents the corresponding $t_k$-year survival probability for this subgroup. In this context, the additional estimating equations can be expressed as:
\[g_k(\bX;
\theta,\Lambda,\pi,\beta) = I(\bZ \in \Omega_k)\left[\exp\left\{-\pi\Lambda(t_k)\exp(\theta^{\rm T}\bZ)\right\}-\beta_k\right]\]
for $k=1,\dots,K$,
where we accommodate the potential heterogeneity of baseline hazard functions by the inclusion of an unknown parameter $\pi$ \citep{huang2016efficient}.

We evaluate the efficacy of the proposed estimators within the following simulation settings. We generated two independent covariates $Z_1 \sim N(0,1)$ and $Z_2 \sim \text{Bernoulli}(0.5)$. The survival time $T$ was generated using the proportional hazards model described in equation (\ref{equ:harzard_func}), with $\boldsymbol{Z}=\left(Z_1, Z_2, Z_1 Z_2\right)^{\mathrm{T}}$, $\theta=(-0.5,1,-0.5)^{\mathrm{T}}$, and $\Lambda(t)=t^2$. To introduce censoring, we generated censoring times from a uniform distribution on the interval [0, 2.52], aiming to achieve a censoring rate of $30 \%$. We first consider the auxiliary survival information that is consistent with the original individual-level data. The auxiliary information is specifically concerned with survival probabilities at time points $t_1=t_2=0.5$ for two subgroups defined by $\Omega_1=\left\{Z_1 \leq 0, Z_2=0\right\}$ and $\Omega_2=\left\{Z_1>0, Z_2=0\right\}$.

\begin{table}[p]
\small
\centering\caption{Simulation results for the proportional hazards model with auxiliary survival probabilities with variability in our paper}
\label{tab:surv1}
\begin{tabular}{|c|c|c|c|c|c|c|c|c|c|c|c|c|c|}
\hline
& & \multicolumn{2}{c}{ MLE } && \multicolumn{2}{c}{ Proposed $(\pi=1)$} && \multicolumn{6}{|c|}{ Proposed ( $\pi$ estimated) } \\\cline{3-4}\cline{5-9}\cline{10-14}
$(n,N, \pi)$ & & Bias & SE & Bias & SE & SEE & CP & RE & Bias & SE & SEE & CP & RE \\
$(100,500,1)$ & $\theta_1$ & -0.021 & 0.218 & -0.024 & 0.167 & 0.154 & 0.929 & 1.7 & -0.025 & 0.167 & 0.152 & 0.928 & 1.7 \\
 &$\theta_2$ & 0.033 & 0.283 & 0.020 & 0.234 & 0.222 & 0.937 & 1.5 & 0.027 & 0.279 & 0.262 & 0.938 & 1.0 \\
 & $\theta_3$ & -0.008 & 0.294 & 0.001 & 0.259 & 0.237 & 0.929 & 1.3 & -0.003 & 0.263 & 0.238 & 0.925 &1.2 \\
& $\pi$ & & & & & & & & 0.064 & 0.371 & 0.351 & 0.930 & \\
$(100,1000,1)$ & $\theta_1$ & -0.024 & 0.215 & -0.019 & 0.143 & 0.135 & 0.940 & 2.3 & -0.017 & 0.142 & 0.132 & 0.938 & 2.3 \\
& $\theta_2$ & 0.033 & 0.284 & 0.005 & 0.223 & 0.213 & 0.938 & 1.6 & 0.020 & 0.278 & 0.261 & 0.937 & 1.0 \\
& $\theta_3$ & -0.003 & 0.291 & 0.003 & 0.244 & 0.227 & 0.935 & 1.4 & -0.007 & 0.248 & 0.228 & 0.933 & 1.4 \\
 & $\pi$ & & & & & & & & 0.066 & 0.333 & 0.322 & 0.938 & \\
$(500,500,1)$ & $\theta_1$ & -0.004 & 0.086 & -0.007 & 0.082 & 0.079 & 0.940 & 1.1 & -0.008 & 0.081 & 0.079 & 0.941 & 1.1 \\
& $\theta_2$ & 0.008 & 0.117 & 0.014 & 0.110 & 0.107 & 0.937 & 1.1 & 0.008 & 0.117 & 0.115 & 0.944 & 1.0 \\
& $\theta_3$ & -0.002 & 0.116 & 0.000 & 0.113 & 0.111 & 0.943 & 1.1 & 0.003 & 0.113 & 0.111 & 0.943 & 1.1 \\
& $\pi$ & & & & & & & & -0.006 & 0.209 & 0.206 & 0.935 & \\
$(500,1000,1)$ & $\theta_1$ & -0.005 & 0.086 & -0.007 & 0.077 & 0.074 & 0.942 & 1.2 & -0.008 & 0.076 & 0.074 & 0.945 & 1.3 \\
& $\theta_2$ & 0.005 & 0.118 & 0.008 & 0.104 & 0.103 & 0.943 & 1.3 & 0.005 & 0.117 & 0.115 & 0.944 & 1.0 \\
 & $\theta_3$ & -0.001 & 0.118 & 0.000 & 0.111 & 0.108 & 0.943 & 1.1 & 0.001 & 0.112 & 0.109 & 0.941 & 1.1 \\
 & $\pi$ & & & & & & & & 0.004 & 0.171 & 0.169 & 0.941 & \\
\hline
\end{tabular}

\end{table}

Table \ref{tab:surv1} presents a summary of the results obtained from our analysis, considering sample sizes of $n=100$ and $n=500$ for the internal study, as well as external sample sizes of $N=500$ and $N=1000$. Specifically, we compare the performance of the maximum likelihood estimator (MLE) against our proposed estimators. When $\pi$ is known to be 1, our proposed estimator demonstrates substantial efficiency gains over the initial MLE, particularly with respect to $\theta_1$. When $\pi$ is estimated, the proposed estimator still exhibits efficiency gains for $\theta_1$ and $\theta_3$ but is less efficient for $\theta_2$. As the size of the external studies increases, the efficiency of our proposed estimator also improves.

In addition, we observe that when the sample size of the internal study increases, the magnitude of efficiency improvement becomes smaller. This trend is likely attributable to the fact that the variance of the initial MLE decreases with a larger internal sample size, making it more challenging to achieve substantial efficiency gains. Importantly, our proposed variance estimators prove to be accurate, as indicated by the reasonable coverage probability of the $95 \%$ confidence intervals. This underlines the robustness and reliability of our proposed methodology.

\begin{table}[p]
\centering\caption{Simulation results for the proportional hazards model with auxiliary survival probabilities with variability}
\label{tab:surv2}
\begin{tabular}{|c|c|c|c|c|c|c|c|c|c|c|c|c|c|}
\hline
& & \multicolumn{2}{c}{ MLE } && \multicolumn{3}{|c|}{ Proposed ( $\pi$ estimated) } \\\cline{3-4}\cline{5-9}
$(n,N, \pi)$ & & Bias & SE & Bias & SE & SEE & CP & RE \\
$(100,500,1.5)$ & $\theta_1$ & -0.024 & 0.215 & -0.023 & 0.157 & 0.144 & 0.936 & 1.9 \\
& $\theta_2$ & 0.033 & 0.284& 0.024 & 0.279 & 0.262 & 0.937 & 1.0 \\
& $\theta_3$ & -0.003 & 0.291 & -0.001 & 0.256 & 0.234 & 0.927 &1.3 \\
$(100,500+500,1.5+2)$ & $\theta_1$ & -0.024 & 0.215 & -0.021 & 0.130 & 0.121 & 0.939 & 2.7 \\
& $\theta_2$ & 0.033 & 0.284& 0.020 & 0.277 & 0.260 & 0.937 & 1.1 \\
 & $\theta_3$ & -0.003 & 0.291 & -0.002 & 0.243 & 0.222 & 0.929 & 1.4 \\
$(500,500,1.5)$ & $\theta_1$ & -0.005 & 0.086& -0.008 & 0.080 & 0.077 & 0.944 & 1.2 \\
& $\theta_2$ & 0.005 & 0.118 & 0.006 & 0.117 & 0.115 & 0.944 & 1.0 \\
& $\theta_3$ & -0.001 & 0.118  & 0.002 & 0.114 & 0.110 & 0.943 & 1.1 \\
$(500,500+500,1.5+2)$ & $\theta_1$ & -0.005 & 0.086 & -0.010 & 0.073 & 0.071 & 0.945 & 1.4 \\
& $\theta_2$ & 0.005 & 0.118 & 0.006 & 0.117 & 0.115 & 0.944 & 1.0 \\
& $\theta_3$ & -0.001 & 0.118 & 0.003 & 0.110 & 0.107 & 0.942 & 1.2 \\
\hline
\end{tabular}

\end{table}

Finally, we examine scenarios with multiple external studies, a frequent situation when leveraging auxiliary information from varied external sources.
Specifically, we consider the incorporation of $t_k$-year survival probabilities estimated from two external studies.
It is worth noting that these different external studies may not necessarily align with the original individual-level data due to variations in inclusion or exclusion criteria within clinical studies. 
In this particular case, we consider different values for $\pi$ for the two external studies: $\pi_1 = 1.5$ and $\pi_2 = 2$.
As shown in Table \ref{tab:surv2}, additionally integrating information from the second external study significantly improves the efficiency of our proposed estimator.
This underscores a pivotal strength of our approach: despite potential inconsistencies among multiple external studies, integrating auxiliary data can still enhance our estimation's efficiency.

\section{Application}

We plan to apply the proposed methods to the analysis of clinical trail data in the Women’s Health Initiative (WHI). The WHI is a long-term national health study aiming to investigate into the factors influencing morbidity and mortality among postmenopausal women. Between 1993 and 1998, WHI enrolled 161,808 postmenopausal women, aged 50–79 years, into clinical trials or a prospective observational study across 40 U.S. study sites (\cite{anderson2003implementation}). The data application result is now under review by the WHI P$\&$P committee. We will update the result once it gets approved.

\section{Discussion}
We have introduced a general framework for improving statistical inference by effectively combining external summary information.
This approach is highly adaptable, capable of accommodating a wide range of summary statistics, even when they originate from models that are different from the internal study. 
A particularly noteworthy feature of our method is its effectiveness in addressing scenarios requiring the consideration of external study variances, a challenge often encountered in the context of semiparametric models with infinite-dimensional parameters. Furthermore, we emphasize the simplicity and efficiency of our framework, especially when dealing with multiple external studies. The resulting updated estimator consistently outperforms the initial estimator, highlighting the utility and effectiveness of our approach.

Our proposed framework builds upon the existing model introducted in \cite{gao2023noniterative}, encompassing scenarios where the variability within auxiliary information cannot be overlooked. This expansion of the current methodology holds substantial importance, particularly within the domain of semiparametric models characterized by parameters of infinite dimensionality. Furthermore, our model is readily amenable to incorporate multiple external studies.

Within our framework, we treat summary information $\beta$ and the nuisance parameter $\pi$ as finite parameters, while infinite-dimensional parameters are primarily associated with internal studies. In some circumstances, summary information from the external studies may involve infinite-dimensional parameters that are different from that in the main study. For instance, the external study may fit a proportional hazards model, such that the baseline hazards function from these external studies is a nuisance parameter. Furthermore, the summary information itself may take an infinite-dimensional form, e.g., the baseline hazard function.
Extending our current methodology to accommodate such settings is challenging both theoretically and computationally.

In some cases, the quality of data from multiple external studies can vary. In such cases, equally incorporation of multiple pieces of auxiliary information may not yield optimal efficiency. A recent study by \cite{gu2023meta} introduced a framework to identify the most relevant external information and reduced the influence of information that is less compatible with the internal data. They proposed a meta-inference framework using an empirical Bayes estimation approach and came up with two weighted estimators to combine the estimates from different external models. We may employ a similar criteria in our framework to determine the weight assigned to each piece of auxiliary information, thus yielding a comprehensive overall estimate. This extension may hold promise for enhancing the efficiency of our results.

\bibliography{Aux_var}

\begin{thebibliography}{}

\bibitem[Anderson et~al., 2003]{anderson2003implementation}
Anderson, G.~L., Manson, J., Wallace, R., Lund, B., Hall, D., Davis, S., Shumaker, S., Wang, C.-Y., Stein, E., and Prentice, R.~L. (2003).
\newblock Implementation of the women's health initiative study design.
\newblock {\em Annals of epidemiology}, 13(9):S5--S17.

\bibitem[Boonstra and Barbaro, 2020]{boonstra2020incorporating}
Boonstra, P.~S. and Barbaro, R.~P. (2020).
\newblock Incorporating historical models with adaptive bayesian updates.
\newblock {\em Biostatistics}, 21(2):e47--e64.

\bibitem[Chatterjee et~al., 2016]{chatterjee2016constrained}
Chatterjee, N., Chen, Y.-H., Maas, P., and Carroll, R.~J. (2016).
\newblock Constrained maximum likelihood estimation for model calibration using summary-level information from external big data sources.
\newblock {\em Journal of the American Statistical Association}, 111(513):107--117.

\bibitem[Cheng et~al., 2019]{cheng2019informing}
Cheng, W., Taylor, J.~M., Gu, T., Tomlins, S.~A., and Mukherjee, B. (2019).
\newblock Informing a risk prediction model for binary outcomes with external coefficient information.
\newblock {\em Journal of the Royal Statistical Society Series C: Applied Statistics}, 68(1):121--139.

\bibitem[Gao and Chan, 2023]{gao2023noniterative}
Gao, F. and Chan, K. (2023).
\newblock Noniterative adjustment to regression estimators with population-based auxiliary information for semiparametric models.
\newblock {\em Biometrics}, 79(1):140--150.

\bibitem[Gu et~al., 2023]{gu2023meta}
Gu, T., Taylor, J.~M., and Mukherjee, B. (2023).
\newblock A meta-inference framework to integrate multiple external models into a current study.
\newblock {\em Biostatistics}, 24(2):406--424.

\bibitem[Han et~al., 2022]{han2022semiparametric}
Han, B., Van~Keilegom, I., and Wang, X. (2022).
\newblock Semiparametric estimation of the nonmixture cure model with auxiliary survival information.
\newblock {\em Biometrics}, 78(2):448--459.

\bibitem[Han and Lawless, 2019]{han2019empirical}
Han, P. and Lawless, J.~F. (2019).
\newblock Empirical likelihood estimation using auxiliary summary information with different covariate distributions.
\newblock {\em Statistica Sinica}, 29(3):1321--1342.

\bibitem[Huang and Qin, 2020]{huang2020unified}
Huang, C.-Y. and Qin, J. (2020).
\newblock A unified approach for synthesizing population-level covariate effect information in semiparametric estimation with survival data.
\newblock {\em Statistics in medicine}, 39(10):1573--1590.

\bibitem[Huang et~al., 2016]{huang2016efficient}
Huang, C.-Y., Qin, J., and Tsai, H.-T. (2016).
\newblock Efficient estimation of the cox model with auxiliary subgroup survival information.
\newblock {\em Journal of the American Statistical Association}, 111(514):787--799.

\bibitem[Kundu et~al., 2019]{kundu2019generalized}
Kundu, P., Tang, R., and Chatterjee, N. (2019).
\newblock Generalized meta-analysis for multiple regression models across studies with disparate covariate information.
\newblock {\em Biometrika}, 106(3):567--585.

\bibitem[Qin, 2000]{qin2000miscellanea}
Qin, J. (2000).
\newblock Miscellanea. combining parametric and empirical likelihoods.
\newblock {\em Biometrika}, 87(2):484--490.

\bibitem[White, 1982]{white1982maximum}
White, H. (1982).
\newblock Maximum likelihood estimation of misspecified models.
\newblock {\em Econometrica: Journal of the econometric society}, pages 1--25.

\bibitem[Zhang et~al., 2020]{zhang2020generalized}
Zhang, H., Deng, L., Schiffman, M., Qin, J., and Yu, K. (2020).
\newblock Generalized integration model for improved statistical inference by leveraging external summary data.
\newblock {\em Biometrika}, 107(3):689--703.

\bibitem[Zhang et~al., 2021]{zhang2021integrative}
Zhang, H., Deng, L., Wheeler, W., Qin, J., and Yu, K. (2021).
\newblock Integrative analysis of multiple case-control studies.
\newblock {\em Biometrics}.

\end{thebibliography}

\end{document}


\title{Supporting Information for: Integration of Summary Information from External Studies for Semiparametric Models}

\author{Jianxuan Zang, K.C.G. Chan, Fei Gao}

\maketitle
\section{Web Appendix A}

\subsection{Derivations of Asymptotic Distributions} \label{append:A.1}
Let $\dbm_\btheta(\btheta,\eta)$ and $\dbm_{\btheta\btheta}(\btheta,\eta)$ be the first and second derivative of $m(\bX;\btheta,\eta)$ with respect to $\btheta$. Let $\dbg_\btheta(\btheta,\eta,\bpi,\bbeta)$,$\dbg_\bpi(\btheta,\eta,\bpi,\bbeta)$, and $\dbg_\bbeta(\btheta,\eta,\bpi,\bbeta)$ as the derivatives of $\bg(\bX;\btheta,\eta,\bpi,\bbeta)$ with respect to $\theta$, $\bpi$, and $\bbeta$. 
In addition, let $\dm_\eta(\btheta,\eta)(h)$ and $\dbg_\eta(\btheta,\eta,\bpi,\bbeta)(h)$ be the derivative of $m(\bX;\btheta,\eta)$ and $\bg(\bX;\btheta,\eta,\bpi,\tbbeta)$ with respect to $\eta$ along the direction indexed by $h\in L_2(\mathP)$, respectively.

We require the following regularity conditions.

{\it Condition 1}. The initial estimator satisfies $\|\hbtheta -\btheta_0\| = o_\mathP(1)$, $\|\hbpi -\bpi_0\| = o_\mathP(1)$, and $\|\heta - \eta_0\|_{\calH} = O_P(n^{-c_1})$ for some $c_1>0$, where $\|\cdot\|$ is the Euclidean norm and $\|\cdot\|_\calH$ denotes the supremum norm on the support of $\eta$.
In addition, the empirical likelihood estimator satisfies $\|\tbtheta -\btheta_0\| = o_\mathP(1)$, $\|\tbpi -\bpi_0\| = o_\mathP(1)$, and $\|\teta - \eta_0\|_{\calH} = O_\mathP(n^{-c_1})$.

{\it Condition 2}. The determinant of the matrix $\{\mathP\bA(\btheta_0,\bpi_0,\eta_0)\}^{-1} \bV\{\mathP\bA(\btheta_0,\bpi_0,\eta_0)\}^{-1}$
 is finite.

{\it Condition 3}. For any $\delta_n\to 0$ and $C>0$,
\[\sup_\calQ\|\mathG_n\{\dbm_\btheta(\btheta,\eta)-\dbm_\btheta(\btheta_0,\eta_0)\}\| = o_\mathP(1),\]
\[\sup_{\calQ,h\in L_2(\mathP)}\|\mathG_n\{\dm_\eta(\btheta,\eta)-\dm_\eta(\btheta_0,\eta_0)\}(h)\| = o_\mathP(1),\]
\[\sup_{\calQ}\|\mathG_n\{\dbg_\btheta(\btheta,\eta,\bpi,\bbeta)-\dbg_\btheta(\btheta_0,\eta_0,\bpi_0,\beta_0)\}\| = o_\mathP(1),\]
\[\sup_{\calQ, h\in L_2(\mathP)}\|\mathG_n\{\dbg_\eta(\btheta,\eta,\bpi,\bbeta)(h)-\dbg_\eta(\btheta_0,\eta_0,\bpi_0,\beta_0)(h)\}\| = o_\mathP(1),\]
\[\sup_{\calQ}\|\mathG_n\{\dbg_\bpi(\btheta,\eta,\bpi,\bbeta)-\dbg_\bpi(\btheta_0,\eta_0,\bpi_0,\beta_0)\}\| = o_\mathP(1),\]
\[\sup_{\calQ}\|\mathG_n\{\dbg_\bbeta(\btheta,\eta,\bpi,\bbeta)-\dbg_\bbeta(\btheta_0,\eta_0,\bpi_0,\beta_0)\}\| = o_\mathP(1),\]
and 
\[\sup_{\calQ}\|\mathG_n\{\bg(\btheta,\eta,\bpi,\bbeta)-\bg(\btheta_0,\eta_0,\bpi_0,\beta_0)\}\| = o_\mathP(1),\]
where $\calQ = \{\|\btheta-\btheta_0\|\le\delta_n, \|\eta - \eta_0\|_{\calH}\le Cn^{-c_1}, \|\bpi-\bpi_0\|\le\delta_n,\|\bbeta-\tbbeta\|\le\delta_n\}$.

{\it Condition 4}. For some $c_2$ satisfying $c_1c_2>1/2$ and for any $(\btheta,\eta,\bpi,\bbeta)\in\calQ$ and $h\in L_2(\mathP)$,
\begin{align*}
&\left\|\mathP\{\dbm_\btheta(\btheta,\eta)-\dbm_\btheta(\btheta_0,\eta_0) - \ddbm_{\btheta\btheta}(\btheta_0,\eta_0)(\btheta-\btheta_0) - \ddbm_{\btheta\eta}(\btheta_0,\eta_0)(\eta-\eta_0)\}\right\|\\
&\qquad= o\left(\left\|\btheta-\btheta_0\right\|\right) + O\left(\left\|\eta-\eta_0\right\|_{\calH}^{c_2}\right),
\end{align*}
\begin{align*}
&\left\|\mathP\{\dm_\eta(\btheta,\eta)(h)-\dm_\eta(\btheta_0,\eta_0)(h) - \ddbm_{\eta\btheta}(\btheta_0,\eta_0)(h)(\btheta-\btheta_0)\right.\\
&\qquad \left.- \ddm_{\eta\eta}(\btheta_0,\eta_0)(h,\eta-\eta_0)\}\right\|= o\left(\left\|\btheta-\btheta_0\right\|\right) + O\left(\left\|\eta-\eta_0\right\|_{\calH}^{c_2}\right),
\end{align*}
and
\begin{align*}
&\left\|\mathP\{\bg(\btheta,\eta,\bpi,\bbeta)-\bg(\btheta_0,\eta_0,\bpi_0,\beta_0) - \dbg_\btheta(\btheta_0,\eta_0,\bpi_0,\beta_0)(\btheta-\btheta_0) - \dbg_\eta(\btheta_0,\eta_0,\bpi_0,\beta_0)(\eta-\eta_0)\}\right.\\
&\qquad \left.- \dbg_\bpi(\btheta_0,\eta_0,\bpi_0,\beta_0)(\bpi-\bpi_0)- \dbg_\bbeta\btheta_0,\eta_0,\bpi_0,\beta_0)(\bbeta-\tbbeta)\right\|\\
&\qquad\qquad= o\left(\left\|\btheta-\btheta_0\right\|+\left\|\bpi-\bpi_0\right\|\right) + O\left(\left\|\eta-\eta_0\right\|_{\calH}^{c_2}\right).
\end{align*}

{\it Condition 5}. The function $\bg(\bX_i;\nu,\eta)$ is twice continuously differentiable in $\mu \in \Omega$, where $\Omega$ is a neighborhood of $\mu_0$, and $E\left\|\bg(\bX_i;\nu,\eta)\right\|^3<\infty$ in $\Omega$.

{\it Condition 6}. The matrix $E\left\{\bg(\bX_i;\nu_0,\eta_0)\bg^{\rm T}(\bX_i;\nu_0,\eta_0)\right\}$ is positive definite at $\mu_0$, and
\begin{align*}
\frac{1}{n}\sum_{i=1}^n \bg(\bX_i;\nu,\eta)\bg^{\rm T}(\bX_i;\nu,\eta)\rightarrow E\left\{\bg(\bX_i;\nu_0,\eta_0)\bg^{\rm T}(\bX_i;\nu_0,\eta_0)\right\}
\end{align*}
in probability, uniformly in $\Omega$ as $n \rightarrow \infty$.

{\it Condition 7}. $E\left\|\frac{\partial\bg(\bX_i;\nu,\eta)}{\partial\mu}\right\| < \infty$ in $\Omega$.

Condition 1 requires the consistency for the initial estimator $(\hbtheta,\heta,\hbpi,\hbbeta)$ and the empirical likelihood estimator $(\tbtheta,\teta,\tbpi,\tbbeta)$ and a certain rate of convergence for $\heta$ and $\teta$, which may be established based on the theory on the consistency and rate of convergence on the estimators.
Condition 2 corresponds to the non-singularity of the information operator, and the asymptotic distributions of $\sqrt n (\hbtheta-\btheta_0)$ and $\sqrt n (\tbtheta-\btheta_0)$ are degenerated if this condition is not satisfied.
Condition 3 is on the stochastic equicontinuity of the first derivative functions, which may be established via entropy calculations and certain maximal inequalities, as demonstrated in \cite{van1996m} and \cite{van2000asymptotic}.
Condition 4 requires the smoothness of the estimating equations, which can be checked by a Taylor series expansion for functionals.

The empirical likelihood estimator $(\hbtheta,\heta,\hbpi,\hbbeta,\hbt)$ satisfies
\begin{align*}
&\mathP_n\left\{\dbm_\btheta(\hbtheta,\heta) - \frac{\dbg_\btheta(\hbtheta,\heta,\hbpi,\hbbeta)^{\rm T}\hbt}{1+\hbt^{\rm T}\bg(\hbtheta,\heta,\hbpi,\hbbeta)}\right\} = \bzero,\\
&\mathP_n\left\{\dm_\eta(\hbtheta,\heta)(h) - \frac{\dbg_\eta(\hbtheta,\heta,\hbpi,\hbbeta)(h)^{\rm T}\hbt}{1+\hbt^{\rm T}\bg(\hbtheta,\heta,\hbpi,\hbbeta)}\right\} = \bzero,\\
&\mathP_n\left\{\frac{\dbg_\bpi(\hbtheta,\heta,\hbpi,\hbbeta)^{\rm T}\hbt}{1+\hbt^{\rm
T}\bg(\hbtheta,\heta,\hbpi,\hbbeta)}\right\} = \bzero,\\
&\mathP_n\left\{\frac{\bg(\hbtheta,\heta,\hbpi,\hbbeta)}{1+\hbt^{\rm T}\bg(\hbtheta,\heta,\hbpi,\hbbeta)}\right\} = \bzero,
\end{align*}
and 
\[\mathP_n\left\{\frac{\dbg_\bbeta(\hbtheta,\heta,\hbpi,\hbbeta)^{\rm T}\hbt}{1+\hbt^{\rm T}\bg(\hbtheta,\heta,\hbpi,\hbbeta)}\right\} + \frac{N}{n}\bV^{-1}\left(\hbbeta - \tbbeta\right) = \bzero,\]
Given Conditions 3 and 4, we expand the first equation to find
\begin{align*}
    -\mathP_n \dbm_\btheta(\btheta_0,\eta_0) = &\mathP_n\left\{\dbm_\btheta(\hbtheta,\heta) - 
    \dbm_\btheta(\btheta_0,\eta_0)\right\}
    -\mathP_n\left\{\frac{\dbg_\btheta(\hbtheta,\heta,\hbpi,\hbbeta)^{\rm T}\hbt}{1+\hbt^{\rm T}\bg(\hbtheta,\heta,\hbpi,\hbbeta)}\right\}\\
    =& \mathP\left\{\dbm_\btheta(\hbtheta,\heta) - 
    \dbm_\btheta(\btheta_0,\eta_0)\right\}
    -\mathP\left\{\frac{\dbg_\btheta(\hbtheta,\heta,\hbpi,\hbbeta)^{\rm T}\hbt}{1+\hbt^{\rm T}\bg(\hbtheta,\heta,\hbpi,\hbbeta)}\right\} + o_\mathP(n^{-1/2})\\
    =& \mathP \ddbm_{\btheta\btheta}(\btheta_0,\eta_0)(\hbtheta-\btheta_0)+\mathP\ddbm_{\btheta\eta}(\btheta_0,\eta_0)(\heta-\eta_0) - \mathP \dbg_\btheta(\btheta_0,\eta_0,\bpi_0,\bbeta_0)^{\rm T}(\hbt-\bzero)\\
    &+ o\left(\left\|\hbtheta-\btheta_0\right\|+\left\|\hbpi-\bpi_0\right\|+\left\|\hbbeta-\bbeta_0\right\|+\left\|\hbt-\bzero\right\|\right) + O\left(\left\|\heta-\eta_0\right\|_\calH^{c_2}\right)+o_\mathP(n^{-1/2})
\end{align*}

Similarly, we expand the rest of the equations to obtain 
\begin{align*}
 - \mathP_n \dm_\eta(\btheta_0,\eta_0)(h)=&\mathP \ddbm_{\btheta\eta}(\btheta_0,\eta_0) (\hbtheta - \btheta_0)(h) + \mathP \ddm_{\eta\eta}(\btheta_0,\eta_0) (\heta - \eta_0)(h)\\&- \mathP \dbg_\eta(\btheta_0,\eta_0,\bpi_0,\bbeta_0)(h)^{\rm T} (\hbt - \bzero) \\& o\left(\left\|\hbtheta - \btheta_0\right\|+\left\|\hbpi - \bpi_0\right\|+\left\|\hbbeta - \bbeta_0\right\|+\left\|\hbt-\bzero\right\|\right)\\&+ O\left(\left\|\heta-\eta_0\right\|_\calH^{c_2}\right) + o_\mathP(n^{-1/2}),\\
\bzero =&\mathP\dbg_\bpi(\btheta_0,\eta_0,\bpi_0,\bbeta_0)^{\rm T}(\hbt - \bzero)+ o\left(\left\|\hbtheta - \btheta_0\right\|+\left\|\hbpi - \bpi_0\right\|+\left\|\hbbeta - \bbeta_0\right\|+\left\|\hbt-\bzero\right\|\right)\\
& + O\left(\left\|\heta-\eta_0\right\|_\calH^{c_2}\right)+ o_\mathP(n^{-1/2}),\\
- \mathP_n \bg(\btheta_0,\eta_0,\bpi_0,\bbeta_0) =& \mathP \dbg_\btheta(\btheta_0,\eta_0,\bpi_0,\bbeta_0)^{\rm T}(\hbtheta- \btheta_0) + \mathP \dbg_\eta(\btheta_0,\eta_0,\bpi_0,\bbeta_0)^{\rm T}(h,\heta- \eta_0) \\
&+ \mathP \dbg_\bpi(\btheta_0,\eta_0,\bpi_0,\bbeta_0)^{\rm T}(\hbpi- \bpi_0) + \mathP \dbg_\bbeta(\btheta_0,\eta_0,\bpi_0,\bbeta_0)^{\rm T}(\hbbeta- \bbeta_0)\\
& - \mathP \bg(\btheta_0,\eta_0,\bpi_0,\bbeta_0)^{\otimes2}(\hbt-\bzero) \\&+ o\left(\left\|\hbtheta - \btheta_0\right\|+\left\|\hbpi - \bpi_0\right\|+\left\|\hbbeta - \bbeta_0\right\|+\left\|\hbt-\bzero\right\|\right) \\
&+ O\left(\left\|\heta-\eta_0\right\|_\calH^{c_2}\right)+ o_\mathP(n^{-1/2}),
\end{align*}
and
\begin{align*}
\bzero = &\mathP\dbg_\bbeta(\btheta_0,\eta_0,\bpi_0,\bbeta_0)^{\rm T}(\hbt - \bzero) + \frac{N}{n}\bV^{-1}\left(\hbbeta - \tbbeta\right) \\
&+ o\left(\left\|\hbtheta - \btheta_0\right\|+\left\|\hbpi - \bpi_0\right\|+\left\|\hbbeta - \bbeta_0\right\|+\left\|\hbt-\bzero\right\|\right)+ O\left(\left\|\heta-\eta_0\right\|_\calH^{c_2}\right) + o_\mathP(n^{-1/2}).
\end{align*}
By comparing the rates, we obtain $\hbtheta-\btheta_0 = O_P(n^{-1/2})$, $\hbpi-\bpi_0 = O_P(n^{-1/2})$, $\hbbeta-\bbeta_0 = O_P(n^{-1/2})$, and $\hbt = O_P(n^{-1/2})$, such that
\begin{align}
\begin{pmatrix}-\mathP \ddbm_{\btheta\btheta}&\bzero_{p\times v}& \mathP \dbg_\btheta^{\rm T}&\bzero_{p\times r}&\bzero_{p\times r}\\
\bzero_{v\times p}&\bzero_{v\times v}& \mathP \dbg_\bpi^{\rm T}&\bzero_{v\times r}&\bzero_{v\times r}\\
-\mathP \dbg_\btheta&- \mathP \dbg_\bpi&\mathP \bg^{\otimes2}&- \mathP \dbg_\bbeta&\bzero_{q\times r}\\
\bzero_{r\times p}&\bzero_{r\times v}& \mathP \dbg_\bbeta^{\rm T}&\rho \bV^{-1}&-\rho \bV^{-1}\end{pmatrix}
\begin{pmatrix}\hbtheta-\btheta_0\\\hbpi - \bpi_0\\\hbt-\bzero\\\hbbeta - \bbeta_0\\\tbbeta - \bbeta_0\end{pmatrix} - \begin{pmatrix}\mathP \ddbm_{\btheta\eta}(\heta-\eta_0)\\
\bzero_{v\times1}\\\mathP \dbg_{\eta}(\heta-\eta_0)\\\bzero_{r\times1}\end{pmatrix}
 = \begin{pmatrix}\mathP_n\dbm_\btheta\\\bzero_{v\times1}\\\mathP_n\bg\\\bzero_{r\times1}\end{pmatrix} + o_\mathP(n{^{-1/2}}).\label{equ:para_func}
\end{align}
where the functions are evaluated at the value $(\btheta_0,\eta_0,\bpi_0,\bbeta_0)$.

In addition, we have 
\begin{align}
& -\mathP \ddbm_{\btheta\eta}(h)\left(\hbtheta-\btheta_0\right) + \mathP \dbg_\eta(h)^{\rm T}\left(\hbt-\bzero\right)- \mathP \ddm_{\eta\eta}(\heta-\eta_0, h)= \mathP_n\dm_\eta(h) + o_\mathP(n^{-1/2})\label{equ:eta_func}
\end{align}
for any $h\in L_2(\mathP)$.
Let $\bh_1^*\in L_2(\mathP)^p$ be the direction that satisfies 
\[\mathP \ddm_{\eta\eta}(h,\bh_1^*) = \mathP \ddbm_{\btheta\eta}(h)\] 
for any $h\in L_2(\mathP)$, and $\bh_2^*\in L_2(\mathP)^q$ be the direction that satisfies 
\[\mathP \ddm_{\eta\eta}(h,\bh_2^*) = \mathP \dbg_{\eta}(h)\] 
or any $h\in L_2(\mathP)$.
Then, by replacing $h$ by $\bh_1^*$ and $\bh_2^*$ in equality (\ref{equ:eta_func}) and comparing with equation (\ref{equ:para_func}), we obtain
\begin{align*}
\begin{pmatrix}-\mathP\tbm_{\btheta\btheta} &\bzero_{p\times v}& \mathP \tbg_\btheta^{\rm T}&\bzero_{p\times r}&\bzero_{p\times r}\\
\bzero_{v\times p}&\bzero_{v\times v}& \mathP\dbg_\bpi^{\rm T}&\bzero_{v\times r}&\bzero_{v\times r}\\
-\mathP\tbg_\btheta&- \mathP\dbg_\bpi &\mathP\bG &- \mathP\dbg_\bbeta&\bzero_{q\times r}\\
\bzero_{r\times p}&\bzero_{r\times v}& \mathP\dbg_\bbeta^{\rm T}&\rho \bV^{-1}&-\rho \bV^{-1}\end{pmatrix}
\begin{pmatrix}\hbtheta-\btheta_0\\\hbpi - \bpi_0\\\hbt-\bzero\\\hbbeta - \bbeta_0\\\tbbeta - \bbeta_0\end{pmatrix} = \begin{pmatrix}\mathP_n\tbm_\btheta \\\bzero_{v\times1}\\\mathP_n\tbg\\\bzero_{r\times1}\end{pmatrix}+ o_\mathP(n^{-1/2}),
\end{align*}
where $\tbm_\btheta = \dbm_\btheta - \dm_\eta(\bh_1^*)$, $\tbg = \bg- \dm_\eta(\bh_2^*)$, $\tbm_{\btheta\btheta} = \ddbm_{\btheta\btheta}-\ddbm_{\btheta\eta}(\bh_1^*)$, $\tbg_\btheta = \dbg_\btheta - \dbg_\eta(\bh_1^*) = \dbg_\btheta-\ddbm_{\btheta\eta}(\bh_2^*)$, and $\bG = \bg ^{\otimes2}-\dbg_\eta(\bh_2^*)^{\rm T}$.
Note that this is equivalent to 
\begin{align}
\begin{pmatrix}-\mathP\tbm_{\btheta\btheta} &\bzero_{p\times v}& \mathP \tbg_\btheta^{\rm T}&\bzero_{p\times r}&\bzero_{p\times r}\\
\bzero_{v\times p}&\bzero_{v\times v}& \mathP\dbg_\bpi^{\rm T}&\bzero_{v\times r}&\bzero_{v\times r}\\
-\mathP\tbg_\btheta&- \mathP\dbg_\bpi &\mathP\bG &- \mathP\dbg_\bbeta&\bzero_{q\times r}\\
\bzero_{r\times p}&\bzero_{r\times v}& \mathP\dbg_\bbeta^{\rm T}&\rho \bV^{-1}&-\rho \bV^{-1}\\
\bzero_{r\times p}& \bzero_{r\times v} &  \bzero_{r\times q} & \bzero_{r\times r}& -\bI_{r\times r}\end{pmatrix}
\begin{pmatrix}\hbtheta-\btheta_0\\\hbpi - \bpi_0\\\hbt-\bzero\\\hbbeta - \bbeta_0\\\tbbeta - \bbeta_0\end{pmatrix} = \begin{pmatrix}\mathP_n\tbm_\btheta \\\bzero_{v\times1}\\\mathP_n\tbg\\\bzero_{r\times1}\\
U\end{pmatrix}+ o_\mathP(n^{-1/2})
\label{equ:middle},
\end{align}
The residual term is still $o_\mathP(n^{-1/2})$ since $N$ is of the same order as $n$.

Thus, the asymptotic distribution of the updated estimator $(\hbtheta,\hbpi,\hbbeta)$ is given by:
\begin{align}
\sqrt n\begin{pmatrix}\hbtheta-\btheta_0\\\hbpi - \bpi_0\\\hbbeta - \bbeta_0\end{pmatrix} &= \begin{pmatrix}\bI_{(p+v)\times(p+v)}&\bzero_{(p+v)\times q}&\bzero_{(p+v)\times r}&\bzero_{(p+v)\times r}\\\bzero_{r\times (p+v)}&\bzero_{r\times q}&\bI_{r\times r}&\bzero_{r\times r}\end{pmatrix}\nonumber\\
&\times\bA(\btheta_0,\eta_0,\bpi_0,\bbeta_0)^{-1}\bl(\btheta_0,\eta_0,\bpi_0,\bbeta_0)+ o_\mathP(1),\label{equ:para_func_final}
\end{align}
where 
\[\resizebox{\hsize}{!}{$
\bA(\btheta,\eta,\bpi,\bbeta) = \begin{pmatrix}-\mathP\tbm_{\btheta\btheta}(\bX;\btheta,\eta) &\bzero_{p\times v}& \mathP\tbg_\btheta^{\rm T}(\bX;\btheta,\eta,\bpi,\bbeta)&\bzero_{p\times r}&\bzero_{p\times r}\\
\bzero_{v\times p}&\bzero_{v\times v}&\mathP\dbg_\bpi^{\rm T}(\bX;\btheta,\eta,\bpi,\bbeta)&\bzero_{v\times r}&\bzero_{v\times r}\\
-\mathP\tbg_\btheta(\bX;\btheta,\eta,\bpi,\bbeta)&- \mathP\dbg_\bpi (\bX;\btheta,\eta,\bpi,\bbeta)&\mathP \bG(\bX;\btheta,\eta,\bpi,\bbeta)&- \mathP\dbg_\bbeta&\bzero_{q\times r} \\
\bzero_{r\times p}&\bzero_{r\times v}&\mathP\dbg_\bbeta^{\rm T}&\rho \bV^{-1}&-\rho \bV^{-1}\\
\bzero_{r\times p}& \bzero_{r\times v} &  \bzero_{r\times q} & \bzero_{r\times r}& \bI_{r\times r}\end{pmatrix}$,}\]
\[\bl(\btheta,\eta,\bpi,\bbeta) = \left(\sqrt n\mathP_n \tbm_\btheta(\bX;\btheta,\eta)^{\rm T}, \bzero_{v\times1}^{\rm T},\sqrt n\mathP_n \tbg(\bX;\btheta,\eta,\bpi,\bbeta)^{\rm T},\bzero_{r\times1}^{\rm T}, \sqrt n U^{\rm T}\right)^{\rm T}\]

As for the initial estimator, we expand the derivatives of $m(\bX;\btheta,\eta)$ and $\bg^*(\bX;\tbtheta,\teta,\bpi,\tbbeta)$ to find
\[-\begin{pmatrix}\mathP\ddbm_{\btheta\btheta} &\bzero_{p\times v}&\bzero_{p\times r}\\
\mathP\dbg^*_{\btheta}&\mathP\dbg^*_{\bpi}&\mathP\dbg^*_{\bbeta}\\
\bzero_{r\times p}&\bzero_{r\times v}&\bI_{r\times r}
\end{pmatrix}
\begin{pmatrix}\tbtheta-\btheta_0\\\tbpi - \bpi_0\\ \tbbeta-\bbeta_0\end{pmatrix} - \begin{pmatrix}\mathP\ddbm_{\btheta\eta}(\teta-\eta_0)\\
\mathP\dbg^*_{\eta}(\teta-\eta_0)\\
0
\end{pmatrix} 
= \begin{pmatrix} \mathP_n\dbm_\btheta \\ \mathP_n\bg^*\\U
\end{pmatrix} + o_\mathP(n^{-1/2}),\]
where $\dbg^*_{\btheta}$, $\dbg^*_{\bpi}$, $\dbg^*_{\bbeta}$ and $\dbg^*_{\eta}(h)$ are the derivatives of $\bg^*$ with respect to $\btheta$, $\bpi$, $\bbeta$, and $\eta$ along the direction indexed by $h$, respectively.
In addition, we have 
\[-\mathP\ddbm_{\btheta\eta}(h)\left(\tbtheta-\btheta_0\right) - \mathP\ddm_{\eta\eta}(\teta-\eta_0, h) = \mathP_n \dm_\eta(h) + o_\mathP(n^{-1/2}),\]
for any $h\in L_2(\mathP)$.
By the definitions of $\bh_1^*$ and $\bh_2^*$, we obtain:
\[\begin{pmatrix}-\mathP\tbm_{\btheta\btheta} &\bzero_{p\times v}&\bzero_{p\times r}\\
-\mathP\tbg^*_{\btheta}&-\mathP\dbg^*_{\bpi} & -\mathP\dbg^*_{\bbeta}\\
\bzero_{r\times p}&\bzero_{r\times v}&-\bI_{r\times r}
\end{pmatrix}
\begin{pmatrix}\tbtheta-\btheta_0\\\tbpi - \bpi_0\\
\tbbeta - \bbeta_0\end{pmatrix} = \begin{pmatrix} \mathP_n\tbm_\btheta\\ \mathP_n\tbg^* \\
U\end{pmatrix}+o_\mathP(n^{-1/2}),\]

That is,
\begin{align}
\sqrt n\begin{pmatrix}\tbtheta-\btheta_0\\\tbpi - \bpi_0\\\tbbeta - \bbeta_0\end{pmatrix} =  & 
\begin{pmatrix}-\mathP\tbm_{\btheta\btheta} &\bzero_{p\times v}&\bzero_{p\times r}\\
-\mathP\tbg^*_{\btheta}&-\mathP\dbg^*_{\bpi} &- \mathP\dbg^*_{\bbeta}\\
\bzero_{r\times p}&\bzero_{r\times v}&-\bI_{r\times r}
\end{pmatrix}^{-1}\begin{pmatrix} \sqrt n\mathP_n\tbm_\btheta(\btheta_0,\eta_0)\\ \sqrt n\mathP_n\tbg^*(\btheta_0,\eta_0,\bpi_0,\bbeta_0)
\\\sqrt n U
\end{pmatrix}+o_\mathP(1)\nonumber\\
=&
\begin{pmatrix}-\mathP\tbm_{\btheta\btheta} &\bzero_{p\times v}&\bzero_{p\times r}\\
-\mathP\tbg^*_{\btheta}&-\mathP\dbg^*_{\bpi} & -\mathP\dbg^*_{\bbeta}\\
\bzero_{r\times p}&\bzero_{r\times v}&-\bI_{r\times r}
\end{pmatrix}^{-1}\nonumber\\
&\times \begin{pmatrix}
\bI_{p\times p} & \bzero_{p\times v} & \bzero_{p\times v} & \bzero_{p\times (q-v)}& \bzero_{p\times r}& \bzero_{p\times r} \\
\bzero_{v\times p}& \bzero_{v\times v}& \bI_{v\times v} & \bzero_{v\times (q-v)} & \bzero_{v\times r} & \bzero_{v\times r} \\
\bzero_{r\times p}& \bzero_{r\times v}& \bzero_{r\times v} & \bzero_{r\times (q-v)} & \bzero_{r\times r} & \bI_{r\times r}
\end{pmatrix}\nonumber\\
&\times \bA(\btheta_0,\eta_0,\bpi_0,\bbeta_0)\bA(\btheta_0,\eta_0,\bpi_0,\bbeta_0)^{-1}\bl(\btheta_0,\eta_0,\bpi_0,\bbeta_0) + o_\mathP(1)\nonumber\\
&=\begin{pmatrix}-\mathP\tbm_{\btheta\btheta} &\bzero_{p\times v}&\bzero_{p\times r}\\
-\mathP\tbg^*_{\btheta}&-\mathP\dbg^*_{\bpi} & -\mathP\dbg^*_{\bbeta}\\
\bzero_{r\times p}&\bzero_{r\times v}&-\bI_{r\times r}
\end{pmatrix}^{-1}\nonumber\\
&\times \begin{pmatrix}
-\mathP\tbm_{\btheta\btheta} & \bzero_{p\times v} & \mathP\tbg_\btheta^{\rm T} &  \bzero_{p\times r}& \bzero_{p\times r} \\
-\mathP\tbg^*_{\btheta} & -\mathP\dbg^*_{\bpi} & \mathP\bG^*&-\mathP\dbg^*_{\bbeta}&\bzero_{v\times r} \\
\bzero_{r\times p}& \bzero_{r\times v}& \bzero_{r\times v}  & \bzero_{r\times r} & -\bI_{r\times r}
\end{pmatrix}\nonumber\\
&\times\bA(\btheta_0,\eta_0,\bpi_0,\bbeta_0)^{-1}\bl(\btheta_0,\eta_0,\bpi_0,\bbeta_0) + o_\mathP(1) ,\label{equ:original_final}
\end{align}
where $\bG^* = \begin{pmatrix} \bI_{v\times v} & \bzero_{v\times (q-v)}\end{pmatrix}\bG$.

Taking the different of (\ref{equ:para_func_final}) and (\ref{equ:original_final}), we have

\begin{align}
\begin{pmatrix}\hbtheta-\tbtheta\\\hbpi - \tbpi\\ \hbbeta-\tbbeta\end{pmatrix}  &= \begin{pmatrix}\bI_{(p+v)\times(p+v)}&\bzero_{(p+v)\times q}&\bzero_{(p+v)\times r}&\bzero_{(p+v)\times r}\\\bzero_{r\times (p+v)}&\bzero_{r\times q}&\bI_{r\times r}&\bzero_{r\times r}\end{pmatrix}\bA^{-1}\bl
\nonumber\\&-
\begin{pmatrix}-\mathP\tbm_{\btheta\btheta} &\bzero_{p\times v}&\bzero_{p\times r}\\
-\mathP\tbg^*_{\btheta}&-\mathP\dbg^*_{\bpi} & -\mathP\dbg^*_{\bbeta}\\
\bzero_{r\times p}&\bzero_{r\times v}&-\bI_{r\times r}
\end{pmatrix}^{-1}
\begin{pmatrix}
-\mathP\tbm_{\btheta\btheta} & \bzero_{p\times v} & \mathP\tbg_\btheta^{\rm T} &  \bzero_{p\times r}& \bzero_{p\times r} \\
-\mathP\tbg^*_{\btheta} & -\mathP\dbg^*_{\bpi} & \mathP\bG^*&-\mathP\dbg^*_{\bbeta}&\bzero_{v\times r} \\
\bzero_{r\times p}& \bzero_{r\times v}& \bzero_{r\times v}  & \bzero_{r\times r} & -\bI_{r\times r}
\end{pmatrix}\bA^{-1}\bl+o_\mathP(1)\nonumber\\
&=  \begin{pmatrix}\bI_{(p+v)\times(p+v)}&\bzero_{(p+v)\times q}&\bzero_{(p+v)\times r}&\bzero_{(p+v)\times r}\\\bzero_{r\times (p+v)}&\bzero_{r\times q}&\bI_{r\times r}&\bzero_{r\times r}\end{pmatrix}\bA^{-1}\bl
\nonumber\\&-
\begin{pmatrix}-\left(\mathP\tbm_{\btheta\btheta}\right)^{-1} &\bzero_{p\times v}&\bzero_{p\times r}\\
\left(\mathP\dbg^*_{\bpi}\right)^{-1}\mathP\dbg^*_{\btheta}\left(\mathP\tbm_{\btheta\btheta}\right)^{-1} & - \left(\mathP\dbg^*_{\bpi}\right)^{-1}&\left(\mathP\dbg^*_{\bpi}\right)^{-1}\mathP\dbg^*_{\bbeta} \\
\bzero_{r\times p}&\bzero_{r\times v}&-\bI_{r\times r}
\end{pmatrix}\nonumber\\
&\times \begin{pmatrix}
-\mathP\tbm_{\btheta\btheta} & \bzero_{p\times v} & \mathP\tbg_\btheta^{\rm T} &  \bzero_{p\times r}& \bzero_{p\times r} \\
-\mathP\tbg^*_{\btheta} & -\mathP\dbg^*_{\bpi} & \mathP\bG^*&-\mathP\dbg^*_{\bbeta}&\bzero_{v\times r}\\
\bzero_{r\times p}& \bzero_{r\times v}& \bzero_{r\times v}  & \bzero_{r\times r} & -\bI_{r\times r}
\end{pmatrix}\bA^{-1}\bl+o_\mathP(1)\nonumber\\
&=  \begin{pmatrix}\bI_{(p+v)\times(p+v)}&\bzero_{(p+v)\times q}&\bzero_{(p+v)\times r}&\bzero_{(p+v)\times r}\\\bzero_{r\times (p+v)}&\bzero_{r\times q}&\bI_{r\times r}&\bzero_{r\times r}\end{pmatrix}\bA^{-1}\bl
\nonumber\\
&-\begin{pmatrix}
\bI_{p\times p} & \bzero_{p\times v} &-\left(\mathP\tbm_{\btheta\btheta}\right)^{-1} \mathP\tbg_\btheta^{\rm T} &  \bzero_{p\times r}& \bzero_{p\times r} \\
\bzero_{v\times p} &\bI_{v\times v} & \left(\mathP\dbg^*_{\bpi}\right)^{-1}\bH&\left(\mathP\dbg^*_{\bpi}\right)^{-1}\mathP\dbg^*_{\bbeta}&-\left(\mathP\dbg^*_{\bpi}\right)^{-1}\mathP\dbg^*_{\bbeta}\\
\bzero_{r\times p} &\bzero_{r\times v} & \bzero_{r\times q}&\bzero_{r\times r}&\bI_{r\times r}
\end{pmatrix}\bA^{-1}\bl+o_\mathP(1)\nonumber\\
&= \begin{pmatrix}
\bzero_{p\times p} & \bzero_{p\times v} &\left(\mathP\tbm_{\btheta\btheta}\right)^{-1} \mathP\tbg_\btheta^{\rm T} &  \bzero_{p\times r}& \bzero_{p\times r} \\
\bzero_{v\times p} &\bzero_{v\times v} &- \left(\mathP\dbg^*_{\bpi}\right)^{-1}\bH&-\left(\mathP\dbg^*_{\bpi}\right)^{-1}\mathP\dbg^*_{\bbeta}&\left(\mathP\dbg^*_{\bpi}\right)^{-1}\mathP\dbg^*_{\bbeta}\\
\bzero_{r\times p} &\bzero_{r\times v} &\bzero_{r\times q} &\bI_{r\times r}&-\bI_{r\times r}
\end{pmatrix}\bA^{-1}\bl+o_\mathP(1)\nonumber
\end{align}

where $\bH = \mathP\tbg^*_{\theta}(\mathP\tbm_{\btheta\btheta})^{-1} \mathP\tbg_\btheta^{\rm T}-\mathP\bG^*$.

\subsection{Proof of theorem 1} \label{append:A.2}

Let $\bmu = (\btheta,\bpi,\bbeta,\eta)$. Let $\Omega_n = \left\{\bmu:\left\|\bmu-\bmu_0\right\| \leq n^{-1/3}\right\}\subset\Omega$
for sufficiently large n. To show the maximum value of log-likelihood function is attained at an interior point of $\Omega_n$, we first give an upper bound of log-likelihood function when $\bmu$ is on the surface of $\Omega_n$.

For $\bmu \in \partial\Omega_n$, we denote a unit vector $u = (\btheta_u,\bpi_u,\bbeta_u,\eta_u)$ satisfying $\bmu = \bmu_0 +un^{-1/3}, \eta = \eta_0 +un^{-1/3}$. By Taylor's expansion
\begin{align*}
&\left\{\sum_{i=1}^nm(\bX_i;\btheta,\eta)- \frac{1}{2}N(\tbbeta - \bbeta)^{\rm T}\bV^{-1}(\tbbeta - \bbeta)\right\}
\\&-\left\{\sum_{i=1}^nm(\bX_i;\btheta_0,\eta_0)- \frac{1}{2}N(\tbbeta - \bbeta_0)^{\rm T}\bV^{-1}(\tbbeta - \bbeta_0)\right\}\\
=&(\btheta-\btheta_0)^{\rm T}\left\{\sum_{i=1}^n\frac{\partial m(\bX_i;\btheta_0,\eta_0)}{\partial\btheta}\right\}
+\frac{1}{2}(\btheta-\btheta_0)^{\rm T}\left\{\sum_{i=1}^n\frac{\partial^2 m(\bX_i;\btheta_0,\eta_0)}{\partial\btheta\partial\btheta^{\rm T}}\right\}(\btheta-\btheta_0)\\
&+(\eta-\eta_0)^{\rm T}\left\{\sum_{i=1}^n\frac{\partial m(\bX_i;\btheta_0,\eta_0)}{\partial\eta}\right\}
+\frac{1}{2}(\eta-\eta_0)^{\rm T}\left\{\sum_{i=1}^n\frac{\partial^2 m(\bX_i;\btheta_0,\eta_0)}{\partial\eta\partial\eta^{\rm T}}\right\}(\eta-\eta_0)\\
&+ N(\bbeta-\bbeta_0)^{\rm T}\bV^{-1}(\tbbeta-\bbeta_0)-\frac{N}{2}(\bbeta-\bbeta_0)^{\rm T}\bV^{-1}(\bbeta-\bbeta_0)+o_p(n^{1/3})\\
\leq & \left\|\btheta-\btheta_0\right\|\left\|\sum_{i=1}^n\frac{\partial m(\bX_i;\btheta_0,\eta_0)}{\partial\btheta}\right\| 
- \frac{n}{2}(\btheta-\btheta_0)^{\rm T}\left\{E\left[-\sum_{i=1}^n\frac{\partial^2 m(\bX_i;\btheta_0,\eta_0)}{\partial\btheta\partial\btheta^{\rm T}}\right]\right\}(\btheta-\btheta_0)\\
&+\left\|\eta-\eta_0\right|\left\|\sum_{i=1}^n\frac{\partial m(\bX_i;\btheta_0,\eta_0)}{\partial\eta}\right\|
-\frac{n}{2}(\eta-\eta_0)^{\rm T}\left\{E\left[-\sum_{i=1}^n\frac{\partial^2 m(\bX_i;\btheta_0,\eta_0)}{\partial\eta\partial\eta^{\rm T}}\right]\right\}(\eta-\eta_0)\\
&+\left\|\bbeta-\bbeta_0\right\|\left\|N\bV^{-1}(\tbbeta-\bbeta_0)\right\| - \frac{N}{2}(\bbeta-\bbeta_0)^{\rm T}\bV^{-1}(\bbeta-\bbeta_0)+o_p(n^{1/3})\\
=& O_p(n^{-1/3}\left\|\btheta_u\right\|n^{1/2}(\log\log n)^{1/2})+O_p(n^{-1/3}\left\|\eta_u\right\|n^{1/2}(\log\log n)^{1/2})\\
&- O_p(n^{1/3}\left\|\btheta_u\right\|^2) -O_p(n^{1/3}\left\|\eta_u\right\|^2)
-O_p(n^{1/3}\left\|\bbeta_u\right\|^2)+o_p(n^{1/3})\\
=& -O_p(n^{1/3})
\end{align*}

$\displaystyle\left\|\sum_{i=1}^n\frac{\partial m(\bX_i;\btheta,\eta)}{\partial\btheta}\right\| = O_p(n^{1/2}(\log\log n)^{1/2})$ is a direct
result by law of iterated logarithm theorem. 

We denote $\bnu = (\btheta,\bpi,\bbeta)$. For any $\bmu \in \partial\Omega_n,t=t(\bnu,\eta)$ satisfies $\sum_{i=1}^n\frac{\bg(\bX_i;\bnu,\eta)}{1+\bt^{\rm T}\bg(\bX_i;\bnu,\eta)}=\bzero$. Assume $E\left\|\bg(\bX;\bnu,\eta)\right\|^3 < \infty$, then by Taylor's expansion and Condition 6-7 we have
\begin{align*}
    t(\bnu,\eta)=&\left\{\frac{1}{n}\sum_{i=1}^n \bg(\bX_i;\bnu,\eta)\bg^{\rm T}(\bX_i;\bnu,\eta)\right\}^{-1} \left\{\frac{1}{n}\sum_{i=1}^n \bg(\bX_i;\bnu,\eta)\right\}+o_p(n^{-1/3})\\
    =&\left[E\left\{\bg(\bX_i;\bnu_0,\eta_0)\bg^{\rm T}(\bX_i;\bnu_0,\eta_0)\right\}\right]^{-1}\left\{\frac{1}{n}\sum_{i=1}^n \bg(\bX_i;\bnu_0,\eta_0)\right.\\
    &+\left.\frac{1}{n}\sum_{i=1}^n\frac{\partial\bg(\bX_i;\bnu_0)}{\partial\bnu}(\bnu-\bnu_0)+
    \frac{1}{n}\sum_{i=1}^n\frac{\partial\bg(\bX_i;\eta_0)}{\partial\eta}(\eta-\eta_0)\right\}+
    o_p(n^{-1/3})\\
    =&O_p(n^{-1/2}(\log\log n)^{1/2}+n^{-1/3}+n^{-1/3})+o_p(n^{-1/3})\\
    =&O_p(n^{-1/3}).
\end{align*}
Therefore,
\begin{align*}
    -&\sum_{i=1}^n \log\left\{1+\bg^{\rm T}(\bX_i;\bnu,\eta)t(\bnu,\eta)\right\}\\
    =-&\sum_{i=1}^n \bg^{\rm T}(\bX_i;\bnu,\eta)t(\bnu,\eta)+\frac{1}{2}\sum_{i=1}^n\left\{\bg^{\rm T}(\bX_i;\bnu,\eta)t(\bnu,\eta)\right\}^2+
    o_p(n^{1/3})\\
    =-&\frac{n}{2}\left\{\frac{1}{n}\sum_{i=1}^n\bg(\bX_i;\bnu,\eta)\right\}^{\rm T}
    \left\{\frac{1}{n}\sum_{i=1}^n\bg(\bX_i;\bnu,\eta)\bg^{\rm T}(\bX_i;\bnu,\eta)\right\}^{-1}
    \left\{\frac{1}{n}\sum_{i=1}^n\bg(\bX_i;\bnu,\eta)\right\} + o_p(n^{1/3})\\
    =-&\frac{n}{2}\left\{\frac{1}{n}\sum_{i=1}^n\bg(\bX_i;\nu_0,\eta_0)+\frac{1}{n}\sum_{i=1}^n
    \frac{\partial\bg(\bX_i;\nu_0)}{\partial\bnu}(\bnu-\bnu_0)+
    \frac{1}{n}\sum_{i=1}^n \frac{\partial\bg(\bX_i;\eta_0)}{\partial\eta}(\eta-\eta_0)\right\}^{\rm T}\\
    \times& \left\{\frac{1}{n}\sum_{i=1}^n\bg(\bX_i;\bnu_0,\eta_0)\bg^{\rm T}(\bX_i;\bnu_0,\eta_0)\right\}^{-1}\\
    \times& \left\{\frac{1}{n}\sum_{i=1}^n\bg(\bX_i;\bnu_0,\eta_0)+\frac{1}{n}\sum_{i=1}^n
    \frac{\partial\bg(\bX_i;\bnu_0)}{\partial\bnu}(\bnu-\bnu_0)+
    \frac{1}{n}\sum_{i=1}^n \frac{\partial\bg(\bX_i;\eta_0)}{\partial\eta}(\eta-\eta_0)\right\} + o_p(n^{1/3})\\
    =-&\frac{n}{2}\left[O_p\left\{n^{-1/2}(\log\log n)^{1/2}\right\}+\left(E\left\{
    \frac{\partial\bg(\bX_i;\bnu_0)}{\partial\bnu}\right\}+E\left\{
    \frac{\partial\bg(\bX_i;\eta_0)}{\partial\eta}\right\}\right)\bmu n^{-1/3}\right]^{\rm T}\\
    \times&\left[E\left\{\bg(\bX_i;\bnu_0,\eta_0)\bg^{\rm T}(\bX_i;\bnu_0,\eta_0)\right\}\right]^{-1}\\
    \times& \left[O_p\left\{n^{-1/2}(\log\log n)^{1/2}\right\}+\left(E\left\{
    \frac{\partial\bg(\bX_i;\bnu_0)}{\partial\bnu}\right\}+E\left\{
    \frac{\partial\bg(\bX_i;\eta_0)}{\partial\eta}\right\}\right)\bmu n^{-1/3}\right] + o_p(n^{1/3})\\
    \leq-&c_1n^{-1/3}+o_p(n^{1/3})\\
    \leq-&c_1n^{-1/3}/2
\end{align*}
where $c_1 > 0$ and it is the smallest eigenvalue of
\[E\left\{\frac{\partial\bg(\bX_i;\bnu_0)}{\partial\bnu}+\frac{\partial\bg(\bX_i;\eta_0)}{\partial\eta}\right\}^{\rm T}\left[E\left\{\bg(\bX_i;\bnu_0,\eta_0)\bg^{\rm T}(\bX_i;\bnu_0,\eta_0)\right\}\right]^{-1}E\left\{\frac{\partial\bg(\bX_i;\bnu_0)}{\partial\bnu}+\frac{\partial\bg(\bX_i;\eta_0)}{\partial\eta}\right\}\]

Similarly,
\begin{align*}
    -&\sum_{i=1}^n \log\left\{1+\bg^{\rm T}(\bX_i;\bnu_0,\eta_0)t(\nu_0,\eta_0)\right\}\\
    =-&\frac{n}{2}\left\{\frac{1}{n}\sum_{i=1}^n\bg(\bX_i;\bnu_0,\eta_0)\right\}^{\rm T}
    \left\{\frac{1}{n}\sum_{i=1}^n\bg(\bX_i;\bnu_0,\eta_0)\bg^{\rm T}(\bX_i;\bnu_0,\eta_0)\right\}^{-1}
    \left\{\frac{1}{n}\sum_{i=1}^n\bg(\bX_i;\bnu_0,\eta_0)\right\} + o_p(1)\\
    =O_p&(\log\log n)
\end{align*}

Denote \begin{equation}
\bl(\bX_i;\bmu) = \sum_{i=1}^nm(\bX_i;\btheta,\teta)-\log\left\{1+\bt^{\rm T}\bg(\bX_i;\btheta,\eta,\bpi,\tbbeta)\right\}- \frac{N}{2}(\tbbeta - \bbeta)^{\rm T}\bV^{-1}(\tbbeta - \bbeta)\nonumber
\end{equation}
then combining all above results we have, for any $\bmu \in \partial \Omega_n$,
\[\bl(\bmu)<\bl(\bmu_0),(a.s.)\]
This means that $\bl(\bmu)$ attains its maximum at $\hat{\bmu}$ in the interior of $\Omega_n$ and $\hat{\bmu}$ satisfies $\displaystyle \frac{\partial \bl(\hat{\bmu})}{\partial\bmu}=0$.

\subsection{Choice of V and proof of the efficiency}\label{append:A.3}
From (\ref{equ:middle}), we can rewrite the estimate equation of the empirical likelihood estimator as:
\begin{align*}
\begin{pmatrix}\mathP \bG &-\mathP\tbg_\btheta&- \mathP\dbg_\bpi&- \mathP\dbg_\bbeta\\
\mathP\tbg_\btheta^{\rm T}&-\mathP\tbm_{\btheta\btheta}&\bzero_{p\times v}& \bzero_{p\times r}\\
\mathP\dbg_\bpi^{\rm T}&\bzero_{v\times p}&\bzero_{v\times v}&\bzero_{v\times r}\\
\mathP\dbg_\bbeta^{\rm T}&\bzero_{r\times p}&\bzero_{r\times v}&\rho \bV^{-1}\end{pmatrix}
\begin{pmatrix}\hbtheta-\btheta_0\\\hbt-\bzero\\\hbpi - \bpi_0\\\hbbeta - \bbeta_0\end{pmatrix} = \begin{pmatrix}\mathP_n\tbm_\btheta \\\mathP_n\tbg\\\bzero_{v\times1}\\-\rho \bV^{-1}U\end{pmatrix}+ o_\mathP(n^{-1/2})
\end{align*}

Denote 
\[\bN = \begin{pmatrix}\mathP \bG &-\mathP\tbg_\btheta&- \mathP\dbg_\bpi&- \mathP\dbg_\bbeta\\
\mathP\tbg_\btheta^{\rm T}&-\mathP\tbm_{\btheta\btheta}&\bzero_{p\times v}& \bzero_{p\times r}\\
\mathP\dbg_\bpi^{\rm T}&\bzero_{v\times p}&\bzero_{v\times v}&\bzero_{v\times r}\\
\mathP\dbg_\bbeta^{\rm T}&\bzero_{r\times p}&\bzero_{r\times v}&\rho \bV^{-1}\end{pmatrix} = \begin{pmatrix}
    \bN_1&\bN_2\\\bN_3&\bN_4(\bV)
\end{pmatrix}\]
where $\bN_1 =\mathP \bG,\bN_4(\bV) = \begin{pmatrix}
-\mathP\tbm_{\btheta\btheta}&\bzero_{p\times v}& \bzero_{p\times r}\\
\bzero_{v\times p}&\bzero_{v\times v}&\bzero_{v\times r}\\
\bzero_{r\times p}&\bzero_{r\times v}&\rho \bV^{-1}\end{pmatrix}$.

Write $\bM(\bV) = (\bN_4(\bV)-\bN_3\bN_1^{-1}\bN_2)^{-1}$, we assume that $\mathP \tbm_{\btheta}^{\otimes2} = -\mathP\tbm_{\btheta\btheta}$ which is the regular condition when $\tbm$ is the log-likelihood function. We calculate the covariance of $\bnu = (\btheta,\bpi,\bbeta) $ under the covariance matrix $\bV$ and $\bSigma_0$
\begin{align*}
    \text{Cov}(\sqrt{n}\hat\bnu_\bV,\sqrt{n}\hat\bnu_{\bSigma_0}) &=\bN_V^{-1}\begin{pmatrix}\mathP \tbg^{\otimes2}&\bzero_{q\times p}&\bzero_{q\times v}&\bzero_{q\times r}\\
    \bzero_{p\times q}&\mathP\tbm_\btheta^{\otimes2}&\bzero_{p\times v}
&\bzero_{p\times r}\\
\bzero_{v\times q}&\bzero_{v\times p}& \bzero_{v\times v}&\bzero_{v\times r}\\
\bzero_{r\times q}&\bzero_{r\times p}& \bzero_{r\times v}&\rho\bV^{-1}
\end{pmatrix}\bN_{\bSigma_0}^{-1}\\
&=
\begin{pmatrix}
    *&*\\-\bM(\bV)\bN_3\bN_1^{-1}&\bM(\bV)
\end{pmatrix}
\begin{pmatrix}
    \bN_1&0\\0&\bN_4(\bV)
\end{pmatrix}\begin{pmatrix}
    *&-\bN_1^{-1}\bN_2\bM(\bSigma_0)\\
    *&\bM(\bSigma_0)
\end{pmatrix}\\
&=
\begin{pmatrix}
    *&*\\
    *&\bM(\bSigma_0)
\end{pmatrix}
\end{align*}

Thus, we know that $\text{Cov}(\sqrt{n}\hat\bnu_\bV) \rightarrow \bM(\bV), \text{Cov}(\sqrt{n}\hat\bnu_{\bSigma_0}) \rightarrow \bM(\Sigma_0)$ and $ \text{Cov}(\sqrt{n}\hat\bnu_\bV,\sqrt{n}\hat\bnu_{\bSigma_0}) \rightarrow \bM(\bSigma_0)$. These results imply that \[\text{Cov}(\sqrt{n}\hat\bnu_\bV-\sqrt{n}\hat\bnu_{\bSigma_0}) \rightarrow \bM(\bV)+\bM(\bSigma_0)-2\bM(\bSigma_0)  = \bM(\bV)-\bM(\bSigma_0)\geq0\]
Therefore, our estimator attains its minimum at $\bV = \bSigma_0$.

Moreover, we consider 
\begin{align*}
    \text{Cov}(\hbtheta) &= [(\bN_4-\bN_3\bN_1^{-1}\bN_2)^{-1}]_{\btheta}\\
    &=(\bN_{4\btheta}+A\bN_{3\btheta}\bN_1^{-1}(-\bN_1-[-\bN_1^{-1}+\bN_{3\bgamma}^{-1}\begin{pmatrix}
         \bzero_{v\times v}&\bzero_{v\times r}\\
 \bzero_{r\times v}&\rho\bSigma_0^{-1}
    \end{pmatrix}\bN_{2\bgamma}^{-1}]^{-1}\bN_1^{-1}\bN_{2\btheta})^{-1}
\end{align*}
where $\bN_{2\theta},\bN_{3\theta},\bN_{4\theta}$ are the part of $\bN_2,\bN_3,\bN_4$ corresponding to $\theta$ and $\bN_{2\bgamma},\bN_{3\bgamma},\bN_{4\bgamma}$ are the part of $\bN_2,\bN_3,\bN_4$ corresponding to $\bgamma = (\bpi,\bbeta)$.

Since $\bN_1 =  -\mathP\tbm_{\btheta\btheta}< 0$, then we have $-\bN_1-[-\bN_1^{-1}+\bN_{3\bgamma}^{-1}\begin{pmatrix}
         \bzero_{v\times v}&\bzero_{v\times r}\\
 \bzero_{r\times v}&\rho\bSigma_0^{-1}
    \end{pmatrix}\bN_{2\bgamma}^{-1}] \geq0$
Therefore, $\text{Cov}(\hbtheta) \leq A_{4\btheta}^{-1}$, which is the covariance matrix of the MLE estimator based on the internal data.
\subsection{Incorporating Multiple External Studies}\label{append:A.4}
As for the multiple external studies, write $\bg_3(\bbeta_1,\bbeta_2) = \bbeta_1-\bbeta_2$, then the empirical likelihood estimator maximizes:
\begin{align}
\sum_{i=1}^nm(\bX_i;\btheta,\eta)-\log\left\{1+\bt_1^{\rm T}\bg_1(\bX_i;\btheta,\eta,\bpi_1,\tbbeta_1)+\bt_2^{\rm T}\bg_2(\bX_i;\btheta,\eta,\bpi_2,\tbbeta_2)+\bt_3^{\rm T}\bg_3(\tbbeta_1,\tbbeta_2) \right\}\nonumber\\
-\frac{1}{2}N_1(\tbbeta_1 - \bbeta_1)^{\rm T}\bV_1^{-1}(\tbbeta_1 - \bbeta_1)- \frac{1}{2}N_2(\tbbeta_2 - \bbeta_2)^{\rm T}\bV_2^{-1}(\tbbeta_2 - \bbeta_2),\label{equ:lik2}
\end{align}
with $\bt_1,\bt_2,\bt_3$ satisfies
\[\sum_{i=1}^n\frac{\bg_1(\bX_i;\btheta,\eta,\bpi_1,\tbbeta_1)}{1+\bt_1^{\rm T}\bg_1(\bX_i;\btheta,\eta,\bpi_1,\tbbeta_1)+\bt_2^{\rm T}\bg_2(\bX_i;\btheta,\eta,\bpi_2,\tbbeta_2)+\bt_3^{\rm T}\bg_3(\tbbeta_1,\tbbeta_2)}=\bzero.\]
\[\sum_{i=1}^n\frac{\bg_2(\bX_i;\btheta,\eta,\bpi_2,\tbbeta_2)}{1+\bt_1^{\rm T}\bg_1(\bX_i;\btheta,\eta,\bpi_1,\tbbeta_1)+\bt_2^{\rm T}\bg_2(\bX_i;\btheta,\eta,\bpi_2,\tbbeta_2)+\bt_3^{\rm T}\bg_3(\tbbeta_1,\tbbeta_2)}=\bzero.\]
\[\sum_{i=1}^n\frac{\bg_3(\tbbeta_1,\tbbeta_2)}{1+\bt_1^{\rm T}\bg_1(\bX_i;\btheta,\eta,\bpi_1,\tbbeta_1)+\bt_2^{\rm T}\bg_2(\bX_i;\btheta,\eta,\bpi_2,\tbbeta_2)+\bt_3^{\rm T}\bg_3(\tbbeta_1,\tbbeta_2)}=\bzero.\]

By a similar procedure from above, we can obtain

{\fontsize{8}{10}\selectfont$
\begin{pmatrix}-\mathP\tbm_{\btheta\btheta} &\bzero_{p\times v}&\bzero_{p\times v}& \mathP \tbg_{1\btheta}^{\rm T}&\mathP \tbg_{2\btheta}^{\rm T}&\bzero_{p\times r}&\bzero_{p\times r}&\bzero_{p\times r}&\bzero_{p\times r}&\bzero_{p\times q}\\
\bzero_{v\times p}&\bzero_{v\times v}&\bzero_{v\times v}& \mathP\dbg_{1\bpi_1}^{\rm T}&\bzero_{v\times q}&\bzero_{v\times r}&\bzero_{v\times r}&\bzero_{v\times r}&\bzero_{v\times r}&\bzero_{v\times q}\\
\bzero_{v\times p}&\bzero_{v\times v}&\bzero_{v\times v}&\bzero_{v\times q}& \mathP\dbg_{2\bpi_2}^{\rm T}&\bzero_{v\times r}&\bzero_{v\times r}&\bzero_{v\times r}&\bzero_{v\times r}&\bzero_{v\times q}\\
-\mathP\tbg_{1\btheta}&- \mathP\dbg_{1\bpi_1}&\bzero_{q\times v}&\mathP\bG_1&\mathP\tbg_{12}&- \mathP\dbg_{1\bbeta_1}&\bzero_{q\times r}& \bzero_{q\times r}&\bzero_{q\times r}&\bzero_{q\times q}\\
-\mathP\tbg_{2\btheta}&\bzero_{q\times v}&- \mathP\dbg_{2\bpi_2}&\mathP\tbg_{21}&\mathP\bG_2&\bzero_{q\times r}&- \mathP\dbg_{2\bbeta_2}&\bzero_{q\times r}& \bzero_{q\times r}&\bzero_{q\times q}\\
\bzero_{r\times p}&\bzero_{r\times v}&\bzero_{r\times v}& \mathP\dbg_{1\bbeta_1}^{\rm T}&\bzero_{r\times q}&\rho_1\bV_1^{-1}&\bzero_{r\times r}&-\rho_1 \bV_1^{-1}&\bzero_{r\times r}&1_{r\times q}\\
\bzero_{r\times p}&\bzero_{r\times v}&\bzero_{r\times v}&\bzero_{r\times q}&\mathP\dbg_{2\bbeta_2}^{\rm T}&\bzero_{r\times r}&\rho_2\bV_2^{-1}&\bzero_{r\times r}&-\rho_2 \bV_2^{-1}&-1_{r\times q}\\
\bzero_{r\times p}& \bzero_{r\times v} &\bzero_{r\times v}&\bzero_{r\times q}&\bzero_{r\times q} & \bzero_{r\times r}& \bzero_{r\times r}& -\bI_{r \times r}&\bzero_{r\times r}&\bzero_{r\times q}\\
\bzero_{r\times p}& \bzero_{r\times v} &\bzero_{r\times v}&\bzero_{r\times q}&\bzero_{r\times q} & \bzero_{r\times r}&\bzero_{r\times r}& \bzero_{r\times r}& -\bI_{r \times r}&\bzero_{r\times q}\\
\bzero_{q\times p}&\bzero_{q\times v}&\bzero_{q\times v}& \bzero_{q\times q}& \bzero_{q\times q}& \bzero_{q\times r}&\bzero_{q\times r}& -1_{q\times r}&1_{q\times r}&\mathP\bG_3\end{pmatrix}
\begin{pmatrix}\hbtheta-\btheta_0\\\hbpi_1 - \bpi_1\\\hbpi_2 - \bpi_2\\\hbt_1-\bzero\\\hbt_2-\bzero\\\hbbeta_1 - \tbbeta_1\\\hbbeta_2 - \tbbeta_2\\\hbbeta_1 - \bbeta_1\\\hbbeta_2 - \bbeta_2\\\hbt_3-\bzero\end{pmatrix}
=\begin{pmatrix}\mathP_n\tbm_\btheta \\\bzero_{v\times1}\\\bzero_{v\times1}\\\mathP_n\tbg_1\\\mathP_n\tbg_2\\\bzero_{r\times1}\\\bzero_{r\times1}\\
U_1\\
U_2\\\mathP_n\tbg_3\end{pmatrix}+ o_\mathP(n^{-1/2}).
$}

Then after some straightforward calculation, we have

{\fontsize{9}{11}\selectfont$
\begin{pmatrix}-\mathP\tbm_{\btheta\btheta} &\bzero_{p\times v}&\bzero_{p\times v}& \mathP \tbg_{1\btheta}^{\rm T}&\mathP \tbg_{2\btheta}^{\rm T}&\bzero_{p\times r}&\bzero_{p\times r}&\bzero_{p\times r}&\bzero_{p\times r}\\
\bzero_{v\times p}&\bzero_{v\times v}&\bzero_{v\times v}& \mathP\dbg_{1\bpi_1}^{\rm T}&\bzero_{v\times q}&\bzero_{v\times r}&\bzero_{v\times r}&\bzero_{v\times r}&\bzero_{v\times r}\\
\bzero_{v\times p}&\bzero_{v\times v}&\bzero_{v\times v}&\bzero_{v\times q}& \mathP\dbg_{2\bpi_2}^{\rm T}&\bzero_{v\times r}&\bzero_{v\times r}&\bzero_{v\times r}&\bzero_{v\times r}\\
-\mathP\tbg_{1\btheta}&- \mathP\dbg_{1\bpi_1}&\bzero_{q\times v}&\mathP\bG_1&\mathP\tbg_{12}&- \mathP\dbg_{1\bbeta_1}&\bzero_{q\times r}& \bzero_{q\times r}&\bzero_{q\times r}\\
-\mathP\tbg_{2\btheta}&\bzero_{q\times v}&- \mathP\dbg_{2\bpi_2}&\mathP\tbg_{21}&\mathP\bG_2&\bzero_{q\times r}&- \mathP\dbg_{2\bbeta_2}&\bzero_{q\times r}& \bzero_{q\times r}\\
\bzero_{r\times p}&\bzero_{r\times v}&\bzero_{r\times v}& \mathP\dbg_{1\bbeta_1}^{\rm T}&\bzero_{r\times q}&\rho_1\bV_1^{-1}&\bzero_{r\times r}&-\rho_1 \bV_1^{-1}&\bzero_{r\times r}\\
\bzero_{r\times p}&\bzero_{r\times v}&\bzero_{r\times v}&\bzero_{r\times q}&\mathP\dbg_{2\bbeta_2}^{\rm T}&\bzero_{r\times r}&\rho_2\bV_2^{-1}&\bzero_{r\times r}&-\rho_2 \bV_2^{-1}\\
\bzero_{r\times p}& \bzero_{r\times v} &\bzero_{r\times v}&\bzero_{r\times q}&\bzero_{r\times q} & \bzero_{r\times r}& \bzero_{r\times r}& -\bI_{r \times r}&\bzero_{r\times r}\\
\bzero_{r\times p}& \bzero_{r\times v} &\bzero_{r\times v}&\bzero_{r\times q}&\bzero_{r\times q} & \bzero_{r\times r}&\bzero_{r\times r}& \bzero_{r\times r}& -\bI_{r \times r}\end{pmatrix}
\begin{pmatrix}\hbtheta-\btheta_0\\\hbpi_1 - \bpi_1\\\hbpi_2 - \bpi_2\\\hbt_1-\bzero\\\hbt_2-\bzero\\\hbbeta_1 - \tbbeta_1\\\hbbeta_2 - \tbbeta_2\\\hbbeta_1 - \bbeta_1\\\hbbeta_2 - \bbeta_2\end{pmatrix}=
\begin{pmatrix}\mathP_n\tbm_\btheta \\\bzero_{v\times1}\\\bzero_{v\times1}\\\mathP_n\tbg_1\\\mathP_n\tbg_2\\\bzero_{r\times1}\\\bzero_{r\times1}\\
U_1\\
U_2\end{pmatrix}+o_\mathP(n^{-1/2})$}

This is actually the same result we get when there is no $\bg_3$, which means that even the two external studies provide the same information, we could treat them as if they are for non-overlapping parameters and apply the proposed non-iterative algorithm.

\section{Web Appendix B}\label{append:B}

\textit{Parametric model}

The function $\bg(\bX;\btheta,\bpi,\bbeta)$ for the parametric model is calculated as follows. In the external study, we estimate $(\pi_1,\beta_1)$ by e.g., least-squares, such that we minimize $\sum_{i=n+1}^{n+N}(Y_i - \pi_1-\beta_1Z_{i1})^2$, which means that we $(\tilde\pi_1,\tilde\beta_1)$ is the solution to 
\[\sum_{i=n+1}^{n+N}(Y_i - \pi_1-\beta_1Z_{i1})(1,Z_{i1})^{\rm T}=0.\]
Therefore, $h(Z;Y) = \left((Y - \pi_1-\beta_1Z_{1})(1,Z_{1})^{\rm T},(Y - \pi_2-\beta_2Z_{2})(1,Z_{2})^{\rm T}\right)$.
In addition, we have 
\[f(Y|Z,\theta) = \frac{1}{\sqrt{2\pi}}\exp\left\{-\frac{(Y - \theta_0 - \theta_1Z_1-\theta_2Z_2)^2}{2}\right\}\]
Then we can obtain
\[\bg(\bX;\btheta,\bpi,\bbeta) = \int_Y h(Z;Y)f(Y|Z,\theta)dY =\left(\begin{matrix}
\theta_0 + \theta_1Z_1+\theta_2Z_2- \pi_1-\beta_1Z_{1}\\
\theta_0 + \theta_1Z_1+\theta_2Z_2- \pi_2-\beta_2Z_{2}\\
(\theta_0 + \theta_1Z_1+\theta_2Z_2- \pi_1-\beta_1Z_{1})Z_1 \\
(\theta_0 + \theta_1Z_1+\theta_2Z_2- \pi_2-\beta_2Z_{2})Z_2
\end{matrix}\right)\]

And for the covariance matrix $\bV$, we calculate from the internal study by the sandwich estimator $\bJ^{-1}\bK\bJ^{-1}$ where
\[\bJ= \left(\begin{matrix}
-1&-Z_1&0&0\\
-Z_1&-Z_1^2&0&0\\
0&0&-1&-Z_2\\
0&0&-Z_2&-Z_2^2
\end{matrix}\right)\;\text{and}\;\bK=\text{Cov}\left(\begin{matrix}
Y- \pi_1-\beta_1Z_{1}\\
(Y- \pi_1-\beta_1Z_{1})Z_1 \\
Y- \pi_2-\beta_2Z_{2}\\
(Y- \pi_2-\beta_2Z_{2})Z_2
\end{matrix}\right)\]

\begin{table}[p]
\centering
\caption{Computation time for the parametric model}
\label{tab:web1}
\begin{tabular}{ccccccc}
\toprule
& \multicolumn{3}{c}{GIM} & \multicolumn{3}{c}{Proposed} \\
\cmidrule(r){3-3} \cmidrule(r){6-6}
Setting & & User Time(s) & & & User Time(s) & \\
\midrule
$N=500$ & $\tilde\beta_2$ & 224.78 & & $\tilde\beta_2$ & 2.64 & \\
& $(\tilde\beta_1, \tilde\beta_2)$ & 331.88 & & $(\tilde\beta_1, \tilde\beta_2)$ & 4.95 & \\
$N=1000$ & $\tilde\beta_2$ & 230.84 & & $\tilde\beta_2$ & 3.9 & \\
& $(\tilde\beta_1, \tilde\beta_2)$ & 395.38 & & $(\tilde\beta_1, \tilde\beta_2)$ & 5.38 & \\
$N=2000$ & $\tilde\beta_2$ & 236.69 & & $\tilde\beta_2$ & 4.55 & \\
& $(\tilde\beta_1, \tilde\beta_2)$ & 397.3 & & $(\tilde\beta_1, \tilde\beta_2)$ & 5.65 & \\
\bottomrule
\end{tabular}
\end{table}

\textit{Logistic regression}

We assume that we observe an i.i.d. sample of $\bX \equiv (Y,Z)$. The full model is given by a logistic regression model $\log\left\{\text{pr}(Y=1|Z_1,Z_2)/\text{pr}(Y=0|Z_1,Z_2)\right\} = \theta_0+Z_1\theta_1+Z_2\theta_2+\epsilon$, with $\epsilon\sim N(0,1)$. The internal data was generated by setting $\theta_0=\theta_1=0.1, \theta_2=0.2$ and $Z_1\sim N(0,1),Z_2\sim N(0,2)$ with $\text{Cov}(Z_1,Z_2)=0.6$. On each external dataset, two reduced models $\log\left\{\text{pr}(Y=1|Z_1)/\text{pr}(Y=0|Z_1)\right\}=\pi_1+Z_1\beta_1$ and $\log\left\{\text{pr}(Y=1|Z_2)/\text{pr}(Y=0|Z_2)\right\}=\pi_2+Z_2\beta_2$ were fitted for estimating $(\beta_1,\beta_2)$. We assumed that either $\tilde\beta_2$ or $(\tilde\beta_1,\tilde\beta_2)$ were given as summary information.

We obtain the function $\bg(\bX;\btheta,\bpi,\bbeta)$ as follows. 

Set $\displaystyle y_1 = \frac{\exp(\pi_1+\beta_1Z_1)}{1+\exp(\pi_1+\beta_1Z_1)}, y_2 = \frac{\exp(\pi_2+\beta_2Z_2)}{1+\exp(\pi_2+\beta_2Z_2)},y_h = \frac{\exp(\theta_0+\theta_1Z_1+\theta_2Z_2)}{1+\exp(\theta_0+\theta_1Z_1+\theta_2Z_2)}$
\[\bg(\bX;\btheta,\bpi,\bbeta) =\left(\begin{matrix}
y_1-y_h&
(y_1-y_h)Z_1&
y_2-y_h&
(y_2-y_h)Z_2
\end{matrix}\right)^{\rm T}\]

And for the covariance matrix $\bV$, And for the covariance matrix $\bV$, we calculate from the internal study by the sandwich estimator $\bJ^{-1}\bK\bJ^{-1}$ where
\[\bJ= \left(\begin{matrix}
-y_1(1-y_1)&0&-y_1(1-y_1)Z_1&0\\
0&-y_2(1-y_2)&0&-y_2(1-y_2)Z_2\\
-y_1(1-y_1)Z_1&0&-y_1(1-y_1)Z_1^2&0\\
0&-y_2(1-y_2)Z_2&0&-y_2(1-y_2)Z_2^2
\end{matrix}\right)\;,\;\bK=\text{Cov}\left(\begin{matrix}
Y- y_1\\
Y-y_2 \\
(Y- y_1)Z_1\\
(Y- y_2)Z_2
\end{matrix}\right)\]

Notice that the internal data consist of $n=1000$ samples. The external data consist $N=500,1000,2000$ samples. Empirical bias, standard deviation and 95 percent coverage probabilities are based on 10 000 replications. The results are shown in Table \ref{tab:web2}. The conclusions are similar to that for the parametric model.

\begin{table}[p]
\centering\caption{Simulation Results for Binary outcome}
\label{tab:web2}
\begin{tabular}{cccccccccccccc}

\hline

	&		&	\multicolumn{3}{c}{MLE}			&&	\multicolumn{3}{c}{Proposed (given $\tilde\beta_2$)}	&&\multicolumn{4}{c}{Proposed (given $(\tilde\beta_1,\tilde\beta_2$)}								\\\cline{3-4}\cline{6-9}\cline{11-14}
Setting	&		&	Bias	&	SE 	&&	Bias	&	SE	&	CP&RE&&	Bias	&	SE	&	CP&RE		\\
N=500	&	$\theta_1$	&	0.0007	&	0.0714	&&	0.0007	&	0.0714	&	0.949 	&1.0&& 0.001	&	0.060	&	0.952&1.2	\\
	&	$\theta_2$	&	0.0005	&	0.0512 	&&	0.0002	&	0.0431	&	0.955&1.2 &&	0.0001	&	0.0424	&	0.956	&1.2\\
N=1000	&	$\theta_1$	&	0.0013	&	0.0714  	&&	0.0013	&	0.0714	&	0.953&1.0&&	0.0011    &	0.0525	&	0.955&1.4\\
	&	$\theta_2$	&	0.0004 &	0.0513   	&&	0.0001	&	0.0392	&	0.950&1.3	&&	0.0001	&	0.0372	&	0.959&1.4	\\
N=2000	&	$\theta_1$	&	0.0019 	&	0.0717   &&	0.0019	&	0.0717	&	0.950 &1.0&&	0.0004	&	0.0510	&	0.954&1.4	\\
	&	$\theta_2$	&	0.0005	&	0.0510 	&&	-0.0004	&	0.0339	&0.953 &1.5&&	-0.0002	&	0.0301	&0.965&1.7\\
 
\hline

\end{tabular}
\end{table}

\bibliography{Aux_var}